\documentclass[a4paper,fleqn,usenatbib]{mnras}

\usepackage{newtxtext,newtxmath}

\usepackage[T1]{fontenc}
\usepackage{ae,aecompl}

\usepackage{graphicx}	
\usepackage{amsmath}	

\defcitealias{Tang2019}{T19}
\defcitealias{Tang2021}{T21}
\defcitealias{Roberts-Borsani2016}{RB16}

\title[]{Lyman-alpha spectroscopy of extreme [OIII] emitting galaxies at $\mathbf{z\simeq2-3}$: implications for Ly$\mathbf{\alpha}$ visibility and LyC leakage at $\mathbf{z>6}$}

\author[M. Tang et al.]{
Mengtao Tang$^{1,2}$\thanks{tangmtasua@email.arizona.edu}, 
Daniel P. Stark$^{1}$, 
Jacopo Chevallard$^{3}$, 
St\'{e}phane Charlot$^{3}$, \newauthor
Ryan Endsley$^{1}$ and 
Enrico Congiu$^{4,5}$ 
\\
\\
$^{1}$ Steward Observatory, University of Arizona, 933 N Cherry Ave, Tucson, AZ 85721, USA \\
$^{2}$ Department of Physics and Astronomy, University College London, Gower Street, London WC1E 6BT, UK \\
$^{3}$ Sorbonne Universit\'{e}, UPMC-CNRS, UMR7095, Institut d'Astrophysique de Paris, F-75014, Paris, France \\
$^{4}$ Departamento de Astronom\'{i}a, Universidad de Chile, Camino del Observatorio 1515, Las Condes, Santiago, Chile \\
$^{5}$ Las Campanas Observatory - Carnegie Institution for Science, Colina el Pino, Casilla 601, La Serena, Chile \\
}

\pubyear{2020}

\begin{document}
\label{firstpage}
\pagerange{\pageref{firstpage}--\pageref{lastpage}}
\maketitle

\begin{abstract}

Spectroscopic observations of massive $z>7$ galaxies selected to have extremely large [O~{\small III}]+H$\beta$ equivalent width (EW $\sim1500$~\AA) have recently revealed large Ly$\alpha$ detection rates, in contrast to the weak emission seen in the general population. Why these systems are uniquely visible in Ly$\alpha$ at redshifts where the IGM is likely significantly neutral is not clear. With the goal of better understanding these results, we have begun a campaign with MMT and Magellan to measure Ly$\alpha$ in galaxies with similar [O~{\small III}]+H$\beta$ EWs at $z\simeq2-3$. At these redshifts, the IGM is highly ionized, allowing us to clearly disentangle how the Ly$\alpha$ properties depend on the [O~{\small III}]+H$\beta$ EW. Here we present Ly$\alpha$ EWs of $49$ galaxies at $z=2.2-3.7$ with intense [O~{\small III}]+H$\beta$ line emission (EW $=300-3000$~\AA). Our results demonstrate that strong Ly$\alpha$ emission (EW $>20$~\AA) becomes more common in galaxies with larger [O~{\small III}]+H$\beta$ EW, reflecting a combination of increasingly efficient ionizing photon production and enhanced transmission of Ly$\alpha$. Among the galaxies with the most extreme [O~{\small III}]+H$\beta$ emission (EW $\sim1500$~\AA), we find that strong Ly$\alpha$ emission is not ubiquitous, with only $50$ per cent of our population showing Ly$\alpha$ EW $>20$~\AA. Our data suggest that the range of Ly$\alpha$ strengths is related to the observed ellipticity, with those systems that appear edge-on or elongated having weaker Ly$\alpha$ emission. We use these results to interpret the anomalous Ly$\alpha$ properties seen in $z>7$ galaxies with extreme [O~{\small III}]+H$\beta$ emission and discuss implications for the escape of ionizing radiation from these extreme line emitting galaxies.

\end{abstract}

\begin{keywords}
cosmology: observations - galaxies: evolution - galaxies: formation - galaxies: high-redshift
\end{keywords}




\section{Introduction} \label{sec:introduction}

The reionization of hydrogen in the intergalactic medium (IGM) is 
thought to be driven by the radiation from the first luminous objects, 
including both massive stars and active galactic nuclei 
\citep[e.g.][]{Loeb2001,Robertson2013,Robertson2015,Bouwens2015,Madau2015,Stark2016,Dayal2018}. 
Therefore studying the process of reionization offers key clues to investigate the 
history of cosmic structure formation. Over the last two decades, the timeline 
of reionization has been constrained by several observations. Planck 
measurements of the electron-scattering optical depth of the cosmic microwave 
background implies a mid-point reionization redshift of $z\simeq7.7$ \citep{Planck2018}. 
Meanwhile, observations of the Ly$\alpha$ and Ly$\beta$ forests in high redshift quasar spectra 
suggest that reionization is nearly complete by $z\approx6$ \citep[e.g.][]{Fan2006,McGreer2015}.

Spectroscopic measurements of Ly$\alpha$ emission from star-forming galaxies 
provide a complementary probe of the IGM at $z\gtrsim7$ \citep[e.g.][]{Dijkstra2014}, 
where the population of quasars becomes rare \citep{Fan2001,Manti2017}. 
Because of the resonant scattering of Ly$\alpha$ photons by neutral hydrogen, 
the damping wings of the neutral patches in the IGM should suppress the Ly$\alpha$ emission 
from galaxies in the reionization era \citep[e.g.][]{Miralda-Escude1998,Santos2004,Mesinger2008}, 
decreasing the fraction of sources showing prominent Ly$\alpha$ emission \citep[e.g.][]{Stark2010,Fontana2010} 
and the abundance of narrowband-selected Ly$\alpha$ emitters 
\citep[e.g.][]{Malhotra2004,Hu2010,Ouchi2010,Kashikawa2011,Konno2014,Santos2016,Ota2017}. 
Over the last decade, significant effort has been invested in campaigns to measure 
the equivalent width (EW) distribution of Ly$\alpha$ emission over cosmic time. 
Spectroscopic observations \citep[e.g.][]{Fontana2010,Stark2011,Treu2013,Caruana2014,Pentericci2014,Schenker2014,Tilvi2014,Jung2018,Jung2020,Mason2019} have demonstrated that there is a downturn in the fraction of 
strong Ly$\alpha$ emitting galaxies at $z\gtrsim6.5$ (the so-called Ly$\alpha$ fraction, x$_{\rm{Ly}\alpha}$), 
consistent with expectations for a significantly neutral IGM ($x_{\rm{HI}}\gtrsim0.5$) at $z\sim7$ 
(e.g., \citealt{Mesinger2015,Zheng2017,Mason2018,Mason2019,Hoag2019,Whitler2020}; see \citealt{Ouchi2020}, for a review). 

In the last several years, attention has focused on observations of four of 
the most luminous ($H_{160}=25.0-25.3$) known galaxies at $z=7-9$ 
\citep[][hereafter \citetalias{Roberts-Borsani2016}]{Roberts-Borsani2016}. 
The red {\it Spitzer}/IRAC $[3.6]-[4.5]$ colors of these four systems imply 
extremely large [O~{\small III}]+H$\beta$\footnote{In this paper 
the [O~{\scriptsize III}] in [O~{\scriptsize III}]+H$\beta$ refers to 
[O~{\scriptsize III}]~$\lambda\lambda4959,5007$.} EWs ($\simeq900-2000$~\AA), 
roughly twice the average [O~{\small III}]+H$\beta$ EW at $z\sim7-8$ 
(EW$_{\rm{[OIII]+H}\beta}\simeq670$~\AA; \citealt{Labbe2013,deBarros2019,Endsley2021a}). 
All four galaxies exhibit strong Ly$\alpha$ emission at $z>7$ 
(\citealt{Oesch2015,Zitrin2015}; \citetalias{Roberts-Borsani2016}; \citealt{Stark2017}), 
implying a $100$ per cent detection rate at redshifts where the IGM is expected to be mostly neutral. 
Taken together with two other similar $z>7$ extreme [O~{\small III}] emitters 
in the literature \citep{Ono2012,Finkelstein2013}, these detections imply a Ly$\alpha$ emitter fraction 
that is five times larger than what is seen in the general population at $z\sim7-8$ \citep{Stark2017}. 
Why this population presents such strong Ly$\alpha$ emission is still a matter of debate. 
Some have suggested that these luminous systems trace overdense regions with larger-than-average 
ionized bubbles, boosting the transmission of Ly$\alpha$ through the IGM \citep[e.g.][]{Barkana2004,Endsley2021b}. 
Alternatively, the large rest-frame optical line EWs of these galaxies may indicate 
hard ionizing radiation fields, potentially enhancing both the production efficiency and 
the escape fraction of Ly$\alpha$ photons through the galaxies \citep[e.g.][]{Stark2017}.

One of the challenges of interpreting the emerging body of reionization-era 
spectra stems from limitations in our understanding of the galaxies with 
large [O~{\small III}]+H$\beta$ EWs ($=300-3000$~\AA). While this population is common at $z>6$, 
they are rare among continuum-selected samples at lower redshifts. Fortunately a series of 
observational campaigns have begun to identify large samples of extreme [O~{\small III}] 
emitting galaxies at $z\simeq0$ \citep[e.g.][]{Cardamone2009,Senchyna2017,Yang2017b}, 
$z\simeq1$ \citep[e.g.][]{Atek2011,Amorin2014,Amorin2015,Huang2015}, 
and $z\simeq2-3$ \citep[e.g.][]{vanderWel2011,Maseda2014,Forrest2017}, 
opening the door for detailed spectroscopic studies of galaxies 
with similar properties to those at $z>6$ 
\citep[e.g.][]{Labbe2013,Smit2015,Roberts-Borsani2016,deBarros2019,Endsley2021a}. 
In \citet[][hereafter \citetalias{Tang2019}]{Tang2019}, 
we presented results from a large near-infrared spectroscopic campaign 
targeting rest-frame optical emission lines in $z\simeq 2$ galaxies with 
[O~{\small III}] EW = $300$~\AA\ to $2000$~\AA. The combination of 
dust-corrected H$\alpha$ and far-UV continuum luminosities enabled calculation of 
the ionizing photon production efficiency ($\xi_{\rm{ion}}$), 
defined as the ratio of the production rate of hydrogen ionizing photons ($N_{\rm{ion}}$) 
and the UV luminosity at 1500~\AA\ ($L_{\rm{UV}}$, including nebular and stellar continuum) 
corrected for dust attenuation from the diffuse interstellar medium (ISM). 
As had been shown previously in nearby galaxy samples \citep{Chevallard2018}, 
\citetalias{Tang2019} found that $\xi_{\rm{ion}}$ scales with the [O~{\small III}] EW, 
reaching very large values in the most extreme line emitters. The ionization state and dust 
content of the nebular gas are also found to scale with [O~{\small III}] EW, such that 
the most intense [O~{\small III}] emitters tend to have gas that is both highly ionized and nearly dust-free. 
With efficient ionizing photon production and little dust, we expect that the production and 
escape of Ly$\alpha$ photons should be maximized (per $L_{\rm{UV}}$) in galaxies with 
the largest [O~{\small III}] EW, potentially explaining the anomalous Ly$\alpha$ detection rates 
in the \citetalias{Roberts-Borsani2016} sample at $z>7$. This general picture is supported by 
observations at $z\simeq0-1$ \citep[e.g.][]{Cowie2011,Amorin2015,Yang2017a} which suggest that 
intense [O~{\small III}] emitting galaxies do indeed tend to exhibit large EW Ly$\alpha$ emission. 

The next step is to investigate how the Ly$\alpha$ EW distribution varies 
over the full range of [O~{\small III}] EWs expected in the reionization era, 
targeting galaxies at lower redshifts where the IGM is known to be highly ionized.  
Such a dataset would reveal how factors internal to galaxies impact the emergent Ly$\alpha$ luminosity, 
providing an empirical baseline at high redshift that is independent of IGM attenuation. 
This goal has motivated observations of Ly$\alpha$ emission in $z\simeq2-3$ galaxies 
selected to have intense [O~{\small III}] emission in 3D-HST grism spectra \citep{Momcheva2016}. 
The first results were presented in \citet{Du2020}, based on a survey conducted with Keck/LRIS. 
Surprisingly the data revealed no significant correlation between Ly$\alpha$ and [O~{\small III}] EWs 
for galaxies in the range $100$~\AA\ $\lesssim$ [O~{\small III}]~$\lambda\lambda4959,5007$ EW $\lesssim1000$~\AA. 
In this paper, we focus on extending the Ly$\alpha$ statistics to higher optical line equivalent 
widths\footnote{Note that in \citet{Du2020}, the [O~{\scriptsize III}] EW refers to 
[O~{\scriptsize III}]~$\lambda\lambda4959,5007$ EW while throughout this paper we will 
use [O~{\scriptsize III}]~$\lambda5007$ EW and [O~{\scriptsize III}]+H$\beta$ EW. 
Adopting the theoretical flux ratio $I(5007)/I(4959)=3$, we have 
EW$_{\rm{[OIII]}\lambda5007}=3/4\times\rm{EW}_{\rm{[OIII]}\lambda\lambda4959,5007}$. 
Assuming the typical flux ratio of [O~{\scriptsize III}]~$\lambda5007/\rm{H}\beta=6$ 
measured for extreme emission line galaxies (e.g., \citealt{Maseda2014}; \citetalias{Tang2019}), 
we have EW$_{\rm{[OIII]+H}\beta}=1.5\times\rm{EW}_{\rm{[OIII]}\lambda5007}$.} 
([O~{\small III}]~$\lambda5007$ EW $\gtrsim1000$~\AA, or equivalently EW$_{\rm{[OIII]+H}\beta}\gtrsim1500$~\AA), 
with the aim of better understanding the Ly$\alpha$ detections in the $z>7$ \citetalias{Roberts-Borsani2016} 
galaxies (median EW$_{\rm{[OIII]+H}\beta}\simeq1500$~\AA). 
The results presented in \citet{Du2020} suggest that stronger Ly$\alpha$ 
emission does indeed appear in this more extreme population, 
but samples are still small at high redshift, with only two 
[O~{\small III}]-selected galaxies in the EW$_{\rm{[OIII]}\lambda5007}\gtrsim1000$~\AA\ 
(i.e., EW$_{\rm{[OIII]}\lambda\lambda4959,5007}\gtrsim1333$~\AA) regime. 
Here we present new Ly$\alpha$ measurements for $49$ $z\simeq2-3$ galaxies 
with intense [O~{\small III}] emission, including $11$ with 
[O~{\small III}]~$\lambda5007$ EW $\gtrsim1000$~\AA, enabling a 
factor of five improvement in Ly$\alpha$ statistics for the most 
extreme line emitters. With this statistical baseline in hand, 
we can begin to understand how factors internal to the galaxy 
(i.e., radiation field, transmission through the circumgalactic medium) 
impact the visibility of Ly$\alpha$ in the most intense [O~{\small III}] emitters, 
providing new insight into what is likely to be driving the anomalous Ly$\alpha$ 
detection rates seen in similar systems at $z>7$.

The organization of this paper is as follows. We describe the observations and Ly$\alpha$ spectra in Section \ref{sec:observation}. The Ly$\alpha$ spectroscopic properties of our extreme [O~{\small III}] emitters at $z\simeq2-3$ is presented in Section \ref{sec:result}. We discuss the implications of the results for galaxies in the reionization era in Section \ref{sec:discussion}, and summarize our conclusions in Section \ref{sec:summary}. We adopt a $\Lambda$-dominated, flat universe with $\Omega_{\Lambda}=0.7$, $\Omega_{\rm{M}}=0.3$, and H$_0=70$ km s$^{-1}$ Mpc$^{-1}$. All magnitudes in this paper are quoted in the AB system \citep{Oke1983}, and all equivalent widths are quoted in the rest-frame.


\section{Observations and analysis} \label{sec:observation}

We aim to characterize the Ly$\alpha$ properties of galaxies with extremely large equivalent width optical emission lines. The data were taken from our optical (rest-frame UV) spectroscopic survey of extreme [O~{\small III}] emitters at $z=1.3-3.7$ using the Inamori-Magellan Areal Camera \& Spectrograph (IMACS; \citealt{Dressler2011}) on the Magellan Baade telescope and the Binospec \citep{Fabricant2019} on the MMT telescope. Details of the sample selection and spectroscopic observations of this survey are described in \citet[][hereafter \citetalias{Tang2021}]{Tang2021}. In this section, we briefly summarize the rest-frame UV spectroscopy in Section \ref{sec:spec}, then present the Ly$\alpha$ emission line measurements in Section \ref{sec:line}. 

\subsection{MMT/Binospec and Magellan/IMACS spectroscopy} \label{sec:spec}

The rest-frame UV spectra used in this work are presented in \citetalias{Tang2021}, which follow a large spectroscopic effort to obtain rest-frame optical spectra of extreme [O~{\small III}] emitters at $z=1.3-3.7$ (\citetalias{Tang2019}; Tang et al. in prep). The sample of extreme [O~{\small III}] emitters was identified based on the [O~{\small III}] EWs inferred from {\it HST} grism spectra (at $z=1.3-2.4$; \citetalias{Tang2019}) or the $K$-band flux excess (at $z=3.1-3.7$; Tang et al. in prep). We require the extreme [O~{\small III}] emitters to have rest-frame [O~{\small III}] $\lambda\lambda4959,5007$ EW $\simeq300-2000$~\AA, which are chosen to match the values expected to be common in reionization-era systems. Over three observing runs between 2018 and 2019, we have obtained rest-frame UV spectra for $138$ extreme [O~{\small III}] emitters with Magellan/IMACS and MMT/Binospec, targeting UV metal line emission (C~{\small IV} $\lambda\lambda1548,1550$, O~{\small III}] $\lambda\lambda1661,1666$, C~{\small III}] $\lambda\lambda1907,1909$; \citetalias{Tang2021}) and Ly$\alpha$. The Magellan/IMACS spectra were reduced using the Carnegie Observatories System for MultiObject Spectroscopy\footnote{\url{https://code.obs.carnegiescience.edu/cosmos}} pipeline \citep{Dressler2011,Oemler2017}, and the MMT/Binospec spectra were reduced using the publicly available Binospec data reduction pipeline\footnote{\url{https://bitbucket.org/chil\_sai/binospec}} \citep{Kansky2019}. We performed the slit loss correction following the same procedures in \citetalias{Tang2019}, and the absolute flux calibration using observations of slit stars.

\begin{figure*}
\begin{center}
\includegraphics[width=0.95\linewidth]{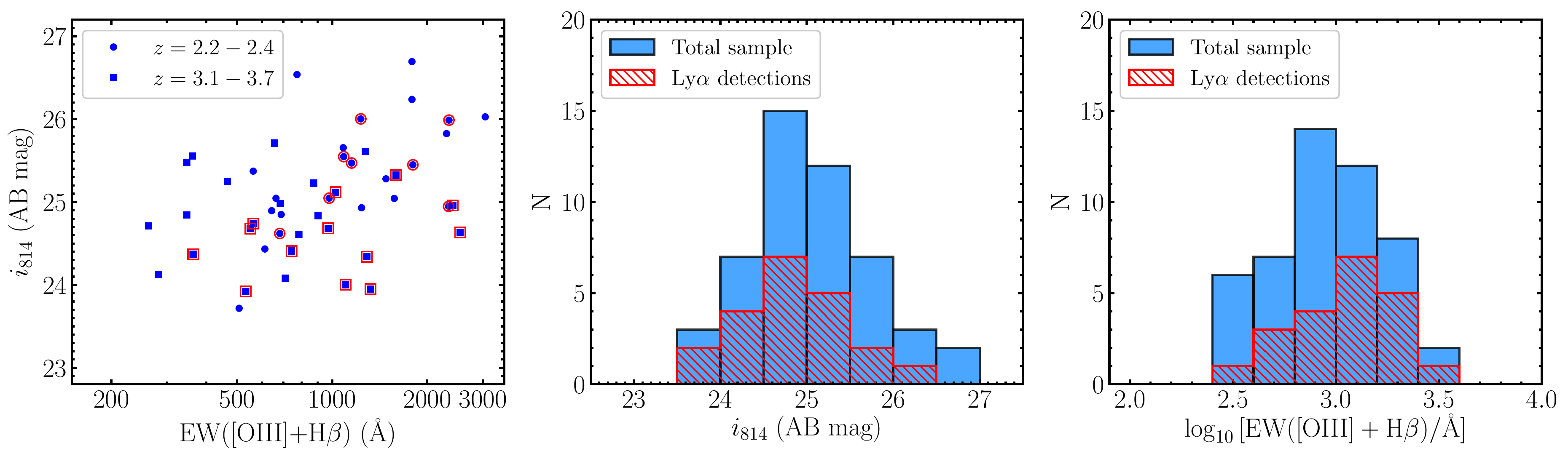}
\caption{{\it HST} F814W magnitude ($i_{814}$) and [O~{\scriptsize III}]+H$\beta$ EW distribution of the $49$ extreme [O~{\scriptsize III}] emitters at $z=2.2-3.7$ in our spectroscopic sample. Left panel: F814W magnitudes versus [O~{\scriptsize III}]+H$\beta$ EWs of the $49$ galaxies (objects at $z=2.2-2.4$ are shown by blue circles, and objects at $3.1-3.7$ are shown by blue squares), sources with Ly$\alpha$ detections are marked by open red circles or squares. Middle panel: F814W magnitude distributions of the total extreme [O~{\scriptsize III}] emitter sample (solid blue) and the subset with Ly$\alpha$ emission line detections (dashed red). Right panel: [O~{\scriptsize III}]+H$\beta$ EW distributions of the total sample (solid blue) and the subset with Ly$\alpha$ detections (dashed red). Our sample spans a range of F814W magnitudes that goes from $24$ to $27$ AB mag, and a wide range of [O~{\scriptsize III}]+H$\beta$ EW ($300-3000$~\AA) which are similar to the values expected at $z>6$.}
\label{fig:mag}
\end{center}
\end{figure*}

\begin{figure*}
\begin{center}
\includegraphics[width=0.95\linewidth]{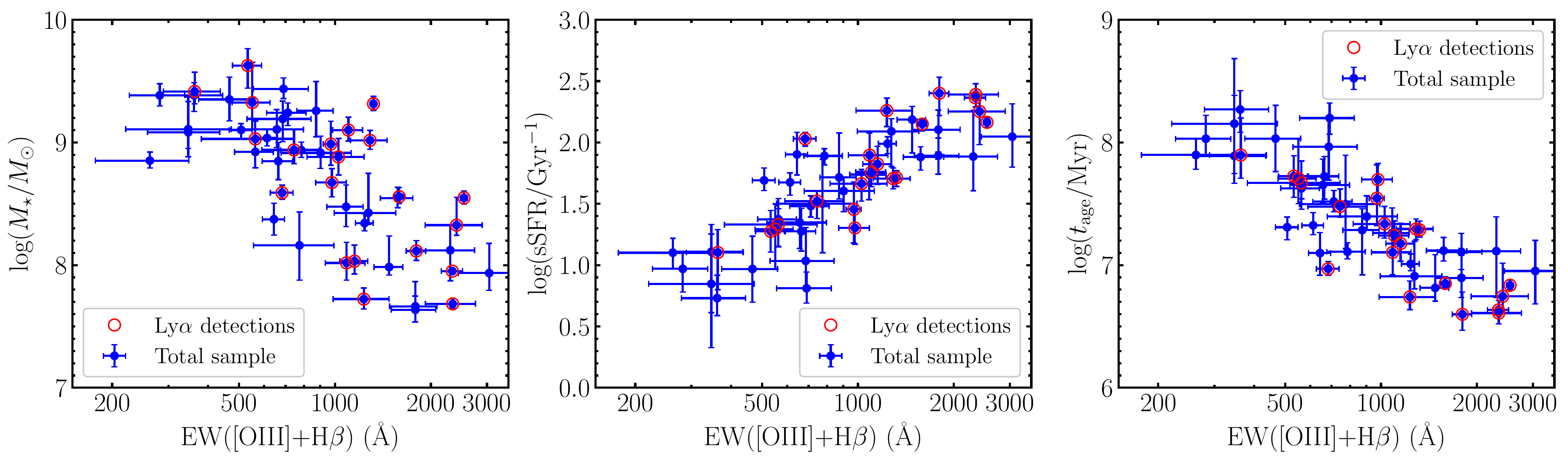}
\caption{Stellar mass (left panel), specific star formation rate (middle), and stellar age (right) as a function of [O~{\scriptsize III}]+H$\beta$ EW for the $49$ extreme [O~{\scriptsize III}] emitters at $z=2.2-3.7$ in our spectroscopic sample. The properties are derived from photoionization modeling using BEAGLE. Galaxies with Ly$\alpha$ emission line measurements (S/N $>3$) are marked by red circles. Systems with larger [O~{\scriptsize III}]+H$\beta$ EWs have lower stellar masses, larger sSFRs, and younger stellar ages.}
\label{fig:ew_star}
\end{center}
\end{figure*}

Our goal is to measure Ly$\alpha$ emission lines in extreme [O~{\small III}] emitters. Due to the wavelength coverage ($\simeq3900-9000$~\AA) of IMACS and Binospec spectra, Ly$\alpha$ is visible for galaxies at $z>2.2$. There are $49$ extreme [O~{\small III}] emitters at $z>2.2$ in our spectroscopic sample. We show the $i_{814}$ magnitude and [O~{\small III}]+H$\beta$ EW distribution of these $49$ sources in Figure \ref{fig:mag}. The median $i_{814}$ magnitude of our sample is $25.0$. We derive the stellar population properties of the $49$ galaxies by fitting the broadband photometry and the available rest-frame optical emission line fluxes using the Bayesian spectral energy distribution (SED) modeling and interpreting tool BEAGLE (version 0.23.0; \citealt{Chevallard2016}). Details of the SED modeling procedures with BEAGLE and the results have been discussed in \citetalias{Tang2021}. In Figure \ref{fig:ew_star}, we show the best-fit stellar masses, specific star formation rates (sSFRs), and stellar ages (assuming constant star formation history) of the $49$ sources. The median [O~{\small III}]+H$\beta$ EW ($901$~\AA) and sSFR ($52$~Gyr$^{-1}$) of our sample are larger than those of typical $z\sim7-8$ galaxies (EW$_{\rm{[OIII]+H}\beta}\sim700$~\AA\ and sSFR $\sim10$~Gyr$^{-1}$; e.g., \citealt{Labbe2013,deBarros2019,Endsley2021a}) since we prioritize targets with the largest EWs ($>1000$~\AA; \citetalias{Tang2021}). However, our sample still spans the full range of EWs expected at $z>6$ (EW$_{\rm{[OIII]+H}\beta}\simeq300-3000$~\AA; e.g., \citealt{Stark2016}), allowing us to investigate how Ly$\alpha$ EW varies over the [O~{\small III}]+H$\beta$ EWs expected in the reionization era.


\subsection{Emission line measurements} \label{sec:line}

We identify Ly$\alpha$ emission lines from the 2D rest-frame UV spectra of the $49$ extreme [O~{\small III}] emitters at $z=2.2-3.7$ by visually inspecting the expected positions of Ly$\alpha$ using the measured redshifts. For $27$ sources in our sample, the redshifts were computed by fitting the [O~{\small III}] $\lambda5007$ emission line from the ground-based \citepalias{Tang2019} or {\it HST} grism-based \citep{Momcheva2016} near-infrared spectra. For the remaining $22$ objects at $z=3.1-3.7$ whose near-infrared spectra are not available, we rely on the photometric redshift measurements from \citet{Skelton2014}. We detected Ly$\alpha$ emission lines with S/N $>3$ in $21$ out of the $49$ extreme [O~{\small III}] emitters at $z=2.2-3.7$ in our spectroscopic sample. For the $22$ objects with photometric redshifts only, Ly$\alpha$ emission was detected in $8$ ($36$ per cent) systems. The Ly$\alpha$ detection rate is higher in the subset with spectroscopic redshifts measured from [O~{\small III}] $\lambda5007$ ($13/27$; $48$ per cent).

\begin{figure*}
\begin{center}
\includegraphics[width=0.95\linewidth]{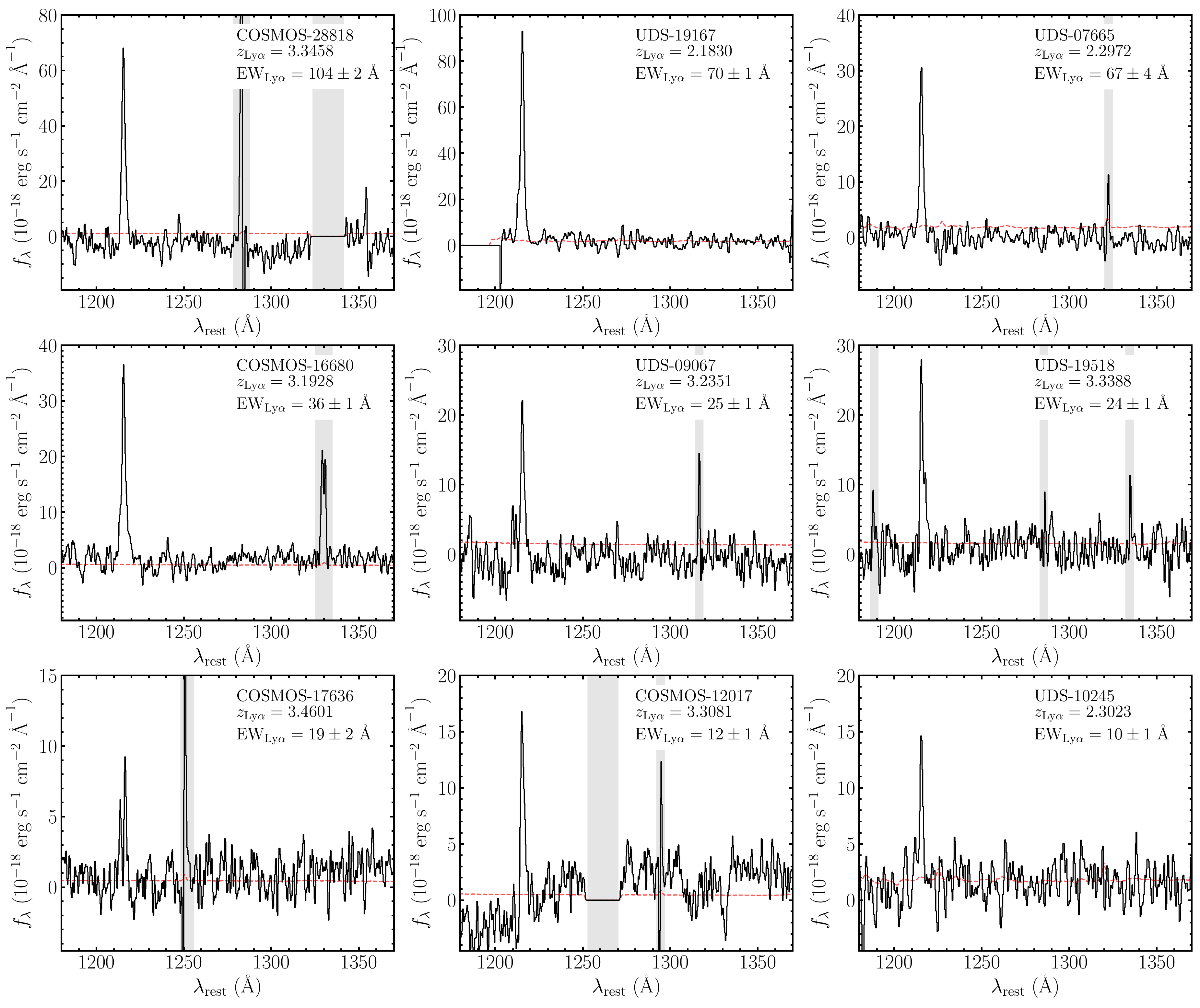}
\caption{Examples of Ly$\alpha$ emission lines presented in the rest-frame UV spectra of $z=1.3-3.7$ extreme [O~{\scriptsize III}] emitters. The black solid lines and red dashed lines represent flux and error, respectively. Detected emission lines are marked by black dotted lines. The grey regions indicate gaps between spectra or wavelength ranges contaminated by sky line residuals.}
\label{fig:lya_spec}
\end{center}
\end{figure*}

Ly$\alpha$ emission line fluxes are determined from the 1D spectra (Figure \ref{fig:lya_spec}), which are extracted from 2D spectra using a boxcar extraction. Twenty of the twenty-one Ly$\alpha$ emitting galaxies show single Ly$\alpha$ emission line features, and the flux was derived by fitting the line profile with a single Gaussian function. The central wavelength recovered from this fit is used to calculate the Ly$\alpha$ redshift. The remaining Ly$\alpha$ emitter (COSMOS-17636) in our sample shows a double-peaked Ly$\alpha$ profile (Figure \ref{fig:lya_spec}), and we fit the emission line with a double-Gaussian function. The line flux is computed by summing the fluxes derived from each single Gaussian profile. For galaxies without S/N $>3$ Ly$\alpha$ emission line measurements, we consider the line as undetected and compute the $3\sigma$ upper limit of the line flux. Using the wavelength boundaries adopted in \citet{Kornei2010}, we derive the $1\sigma$ Ly$\alpha$ flux by integrating the error spectrum in quadrature over rest-frame $1199.9$~\AA\ to $1228.8$~\AA. Since the throughput declines rapidly at the short wavelength end ($<4500$~\AA) of IMACS and Binospec spectrographs, the sensitivity of detecting a Ly$\alpha$ emission line in $z=2.2-2.4$ galaxies is systematically lower than that in $z=3.1-3.7$ galaxies. At $z=2.2-2.4$, the measured Ly$\alpha$ emission line fluxes range from $3.0\times10^{-17}$ ~erg s$^{-1}$ cm$^{-2}$ to $2.2\times10^{-16}$ ~erg s$^{-1}$ cm$^{-2}$, and the median $3\sigma$ flux limit of undetected Ly$\alpha$ is $6.7\times10^{-17}$ ~erg s$^{-1}$ cm$^{-2}$. At $z=3.1-3.7$, the measured Ly$\alpha$ emission line fluxes range from $8.4\times10^{-18}$ ~erg s$^{-1}$ cm$^{-2}$ to $1.7\times10^{-16}$ ~erg s$^{-1}$ cm$^{-2}$, and the median $3\sigma$ flux limit of undetected Ly$\alpha$ is $1.7\times10^{-17}$ ~erg s$^{-1}$ cm$^{-2}$.

We next compute the Ly$\alpha$ emission line EWs. Accurate measurement of Ly$\alpha$ EW is based on both the measurements of Ly$\alpha$ emission line flux and the underlying continuum flux density. Since many of our rest-frame UV spectra do not show high S/N ($>5$) continuum feature near Ly$\alpha$, we take advantage of broadband photometry from \citet{Skelton2014} to estimate the continuum flux density. We consider filters with wavelength coverage between rest-frame $1250$~\AA\ and $2600$~\AA\ (the same wavelength range used to compute UV slope in \citealt{Calzetti1994}), and fit the broadband fluxes with a power-law ($f_{\lambda}\propto\lambda^{\beta}$). From the fitted $f_{\lambda}-\lambda$ relation, we derive the average flux density between $1225$~\AA\ and $1250$~\AA\ \citep{Kornei2010} as the continuum flux density. The Ly$\alpha$ EWs are then computed by dividing the measured Ly$\alpha$ emission line fluxes by the continuum flux densities, ranging from $4$~\AA\ to $142$~\AA\ with a median value of $24$~\AA\ for the $21$ Ly$\alpha$ emitting systems in our sample. Among the $21$ galaxies with Ly$\alpha$ emission line detections, only 8 are at $z=2.2-2.4$ (out of $23$ galaxies observed at this redshift). This is because Ly$\alpha$ is situated at the blue end of the IMACS or Binospec spectra ($\simeq3890-4130$~\AA) where the efficiency declines rapidly ($\simeq30$ per cent of the maximum efficiency). For the $z=2.2-2.4$ galaxies without Ly$\alpha$ detections, the median $3\sigma$ upper limit of Ly$\alpha$ EW is $23$~\AA. On the other hand, half ($13$ out of $26$) of the $z=3.1-3.7$ galaxies are detected with Ly$\alpha$ emission lines, and the median $3\sigma$ upper limit of Ly$\alpha$ EW for those without Ly$\alpha$ detections is $5$~\AA. 

Finally, for a subset ($11$ out of $21$) of Ly$\alpha$ emitting galaxies with O~{\small III}] $\lambda1666$ or [O~{\small III}] $\lambda5007$ emission lines (and hence systemic redshifts) measured from ground-based telescopes, we compute the velocity offset between Ly$\alpha$ and O~{\small III}] or [O~{\small III}]. The Ly$\alpha$ velocity offsets of these $11$ sources are from $-28$ km/s to $766$ km/s, with a median of $164$ km/s. This indicates that the Ly$\alpha$ emission is typically redshifted with respect to oxygen emission lines, but the velocity offsets are lower than the average value ($445$ km/s) of more massive, typical star-forming galaxies at $z\sim2$ \citep{Steidel2010}. In Table \ref{tab:lya}, we summarize the Ly$\alpha$ properties of the $21$ extreme [O~{\small III}] emitters with Ly$\alpha$ emission detections in our spectroscopic sample. 


\begin{table*}
\begin{tabular}{|c|c|c|c|c|c|c|c|c|}
\hline
Target & R.A. & Decl. & $z_{\rm{sys}}$ & $z_{\rm{Ly}\alpha}$ & F$_{\rm{Ly}\alpha}$ & EW$_{\rm{Ly}\alpha}$ & EW$_{\rm{[OIII]+H}\beta}$ & [O~{\scriptsize III}]/[O~{\scriptsize II}] \\
 & (hh:mm:ss) & (dd:mm:ss) & & & ($10^{-18}$ ~erg s$^{-1}$ cm$^{-2}$) & (\AA) & (\AA) & \\
\hline 
\hline
COSMOS-12017 & 10:00:35.387 & +02:18:05.730 & ... & $3.3081$ & $42.21\pm0.98$ & $12.2\pm0.3$ & $532\pm55$ & ... \\ 
COSMOS-16680 & 10:00:48.029 & +02:20:57.824 & $3.1846$ & $3.1921$ & $137.70\pm1.19$ & $36.2\pm0.3$ & $1102\pm118$ & ... \\ 
COSMOS-17636 & 10:00:40.510 & +02:21:32.379 & ... & $3.4601$ & $18.92\pm3.15$ & $11.4\pm1.9$ & $562\pm181$ & ... \\ 
COSMOS-18503 & 10:00:19.083 & +02:22:04.057 & ... & $3.4229$ & $18.90\pm0.74$ & $8.3\pm0.3$ & $744\pm154$ & ... \\ 
COSMOS-19118 & 10:00:25.726 & +02:22:24.225 & ... & $3.4268$ & $8.44\pm0.65$ & $4.4\pm0.3$ & $363\pm73$ & ... \\ 
COSMOS-22402 & 10:00:17.831 & +02:24:26.350 & $2.2751$ & $2.2794$ & $76.59\pm7.91$ & $26.6\pm2.7$ & $682\pm45$ & ... \\ 
COSMOS-27885 & 10:00:36.317 & +02:28:17.384 & ... & $2.2546$ & $79.81\pm11.58$ & $48.1\pm7.0$ & $1087\pm115$ & ... \\ 
COSMOS-28818 & 10:00:40.009 & +02:29:01.853 & ... & $3.3458$ & $168.80\pm2.11$ & $104.1\pm1.3$ & $2409\pm490$ & ... \\ 
COSMOS-31220 & 10:00:40.671 & +02:31:00.551 & ... & $3.4312$ & $20.87\pm0.78$ & $7.6\pm0.3$ & $1288\pm169$ & ... \\ 
UDS-06274 & 02:17:52.310 & -05:15:20.264 & ... & $3.1040$ & $10.00\pm2.61$ & $5.5\pm1.4$ & $550\pm75$ & ... \\ 
UDS-07665 & 02:17:33.781 & -05:15:02.848 & $2.2955$ & $2.2972$ & $72.50\pm4.39$ & $67.7\pm4.1$ & $1800\pm101$ & ... \\ 
UDS-08078 & 02:17:02.741 & -05:14:57.498 & $3.2277$ & $3.2385$ & $18.21\pm2.85$ & $4.9\pm0.8$ & $1321\pm30$ & $9.1\pm0.5$ \\ 
UDS-09067 & 02:17:01.477 & -05:14:45.359 & $3.2288$ & $3.2351$ & $53.72\pm3.07$ & $25.1\pm1.4$ & $2541\pm63$ & $15.3\pm1.5$ \\ 
UDS-10245 & 02:17:22.926 & -05:14:30.628 & $2.2995$ & $2.3023$ & $29.50\pm4.10$ & $11.0\pm1.5$ & $978\pm107$ & $8.1\pm3.0$ \\ 
UDS-10805 & 02:17:23.712 & -05:14:22.974 & $2.2925$ & $2.2934$ & $35.83\pm5.57$ & $27.8\pm4.3$ & $1232\pm245$ & $6.7\pm2.9$ \\ 
UDS-15533 & 02:17:26.075 & -05:13:25.277 & ... & $2.1589$ & $45.71\pm7.19$ & $20.4\pm3.2$ & $1152\pm87$ & ... \\ 
UDS-19167 & 02:17:43.535 & -05:12:43.610 & $2.1833$ & $2.1830$ & $219.00\pm4.67$ & $70.9\pm1.5$ & $2335\pm178$ & $10.4\pm2.6$ \\ 
UDS-19518 & 02:17:19.013 & -05:12:38.390 & $3.3368$ & $3.3388$ & $50.51\pm2.65$ & $24.7\pm1.3$ & $970\pm52$ & $13.3\pm3.8$ \\ 
UDS-21724 & 02:17:20.006 & -05:12:10.624 & $3.2278$ & $3.2291$ & $26.75\pm2.70$ & $23.8\pm2.4$ & $1591\pm51$ & $17.0\pm2.6$ \\ 
UDS-24093 & 02:17:51.205 & -05:11:42.241 & ... & $3.7116$ & $23.99\pm2.45$ & $23.4\pm2.4$ & $1026\pm208$ & ... \\ 
UDS-29766 & 02:17:43.464 & -05:10:33.445 & $2.3023$ & $2.3041$ & $145.70\pm33.55$ & $142.4\pm32.8$ & $2341\pm418$ & $10.1\pm2.9$ \\ 
\hline
\end{tabular}
\caption{List of the $21$ Ly$\alpha$ line emitting galaxies at $z=2.2-3.7$ in our spectroscopic sample, including Ly$\alpha$ fluxes (F$_{\rm{Ly}\alpha}$) and equivalent widths (EW$_{\rm{Ly}\alpha}$). Systemic redshifts ($z_{\rm{sys}}$) are computed by fitting [O~{\scriptsize III}] $\lambda5007$ or O~{\scriptsize III}] $\lambda1666$ emission lines.}
\label{tab:lya}
\end{table*}


\section{Lyman-alpha spectral properties of extreme [O~{\small III}] emitters at $\mathbf{z=1.3-3.7}$} \label{sec:result}

In this section, we use our $z\simeq2-3$ spectroscopic sample to 
quantify the dependence of the Ly$\alpha$ EW on [O~{\small III}]+H$\beta$ EW, 
providing a baseline for interpreting how internal galaxy properties 
impact the production and escape of Ly$\alpha$ in the population of 
extreme line emitters which is common at $z>6$. Work has previously 
shown that the production efficiency of hydrogen ionizing photons 
increases with [O~{\small III}]+H$\beta$ EW (\citealt{Chevallard2018}; \citetalias{Tang2019}), 
suggesting that the most intense [O~{\small III}]+H$\beta$ emitters produce 
more hydrogen ionizing photons relative to $L_{\rm{UV}}$ at $1500$~\AA\ 
than galaxies with lower [O~{\small III}]+H$\beta$ EWs. Since Ly$\alpha$ is 
powered by hydrogen ionizing photons, we expect that the luminosity of 
Ly$\alpha$ relative to $L_{\rm{UV}}$ should also scale with 
[O~{\small III}]+H$\beta$ EW. However the precise scaling of Ly$\alpha$ EW with 
[O~{\small III}]+H$\beta$ EW depends not only on Ly$\alpha$ production but 
also on the escape of Ly$\alpha$ through the ISM and CGM of the galaxy. 
The large specific star formation rates required to produce large 
[O~{\small III}]+H$\beta$ EW could result in extreme feedback conditions that 
maximize the transmission of Ly$\alpha$. How the ISM and CGM modulates 
the escape of Ly$\alpha$ in this class of galaxies is not well 
quantified in a statistical manner, making it difficult to 
interpret the extent to which internal galaxy properties are 
driving the anomalous Ly$\alpha$ seen in galaxies with intense 
[O~{\small III}] emission at $z>7$.

Our rest-frame UV spectroscopic survey of extreme [O~{\small III}] 
emitters allows us 
to make progress in the determination of the Ly$\alpha$ EW distribution 
in galaxies with [O~{\small III}]+H$\beta$ EW $>300$~\AA, building on the 
recent survey presented in \citet{Du2020}.
We consider sources in our sample at $z=2.2-3.7$, the redshift range where 
our optical spectra are able to detect Ly$\alpha$ emission.  
Our current survey contains $49$ extreme [O~{\small III}] emitters
(EW$_{\rm{[OIII]}\lambda\lambda4959,5007}>300$~\AA\ or equivalently
EW$_{\rm{[OIII]+H}\beta}>340$~\AA) with Ly$\alpha$ constraints. 
We have focused our survey on building the sample of galaxies 
with the [O~{\small III}]+H$\beta$ EWs ($>1500$~\AA) exhibited by many of 
the known Ly$\alpha$ detections at $z>7$. We currently have 
obtained Ly$\alpha$ constraints for $11$ objects with 
[O~{\small III}]+H$\beta$ EW $>1500$~\AA. 

\begin{figure}
\begin{center}
\includegraphics[width=\linewidth]{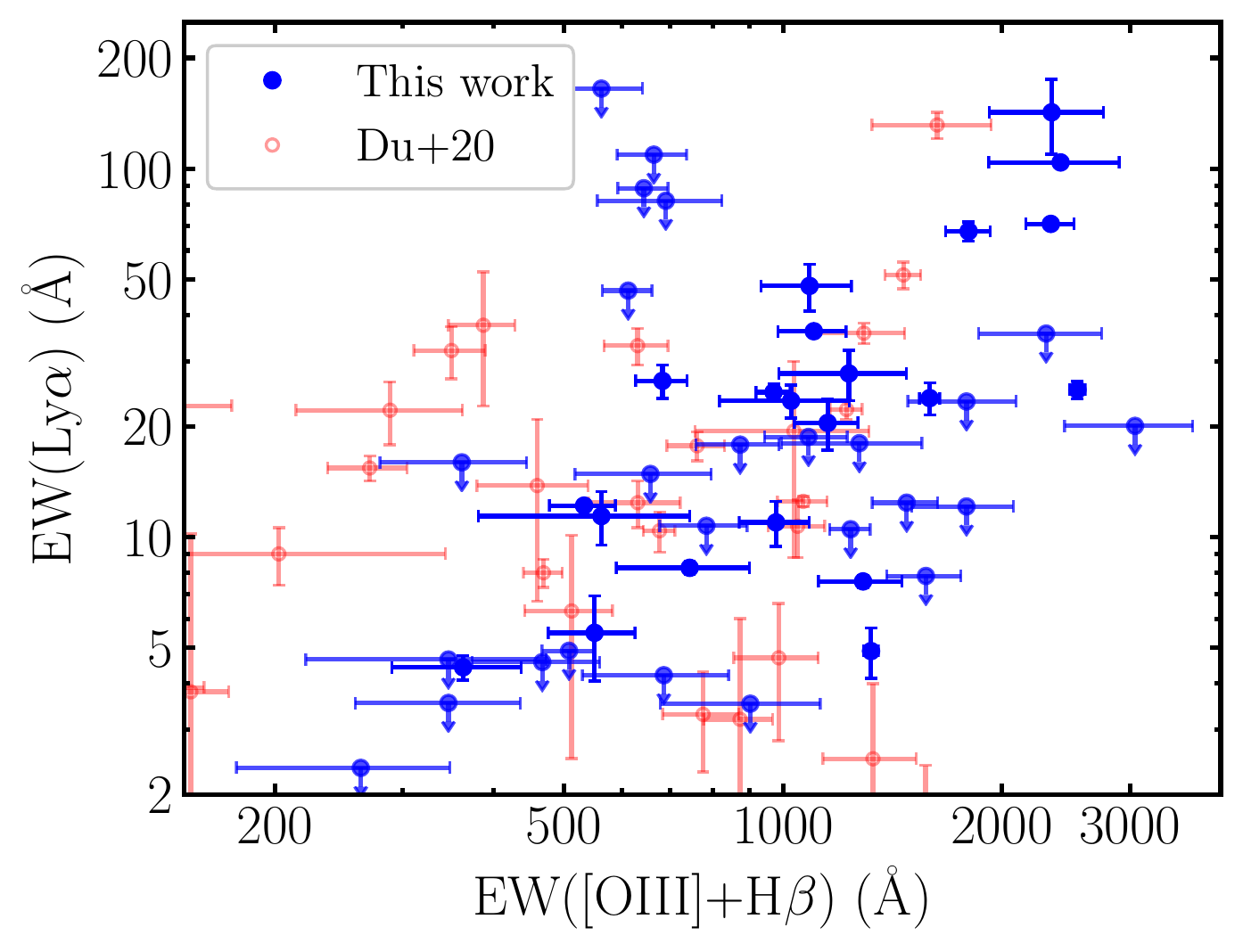}
\caption{Ly$\alpha$ EW as a function of [O~{\scriptsize III}]+H$\beta$ EW for our sample (blue circles) and the sample in \citet{Du2020} (open red circles). Galaxies with large Ly$\alpha$ EW ($>50$~\AA) are absent at [O~{\scriptsize III}]+H$\beta$ EW $<500$~\AA, while this population becomes more common at [O~{\scriptsize III}]+H$\beta$ EW $>1000$~\AA. Also, not all the galaxies with [O~{\scriptsize III}]+H$\beta$ EW $>1000$~\AA\ show strong Ly$\alpha$, with about half of this population show Ly$\alpha$ EW $<10-20$~\AA.}
\label{fig:lyaew_o3hbew}
\end{center}
\end{figure}

In Figure \ref{fig:lyaew_o3hbew}, we present the Ly$\alpha$ EWs
of galaxies in our sample as a function of [O~{\small III}]+H$\beta$ EW. 
We present both detections and non-detections and also include 
the similarly-selected sample from \citet{Du2020}.
Two things are important to take away from the data. First, 
we see an absence of the largest Ly$\alpha$ EWs ($>50$~\AA) 
among the lower [O~{\small III}]+H$\beta$ EWs ($<500$~\AA) in our sample.
Such strong Ly$\alpha$ emitters appear to become more common 
among the most extreme [O~{\small III}]+H$\beta$ (EW $>1000$~\AA), 
as was previously reported in several other studies 
\citep{Yang2017a,Du2020}. At the largest [O~{\small III}]+H$\beta$ EWs 
($>2000$~\AA), we start to see Ly$\alpha$ detections 
with EW$_{\rm{Ly}\alpha}=100-150$~\AA, requiring extremely 
efficient production and transmission. According to the BEAGLE 
photoionization models, these galaxies are dominated by light from 
extremely young stellar populations ($<10$ Myr), with low 
metallicities ($\simeq0.1-0.2\ Z_\odot$) and large ionization 
parameters ($\log{U}=-2.0$ to $-1.5$), as expected for a galaxy 
that has recently experienced a significant upturn in its star formation.  

\begin{figure}
\begin{center}
\includegraphics[width=\linewidth]{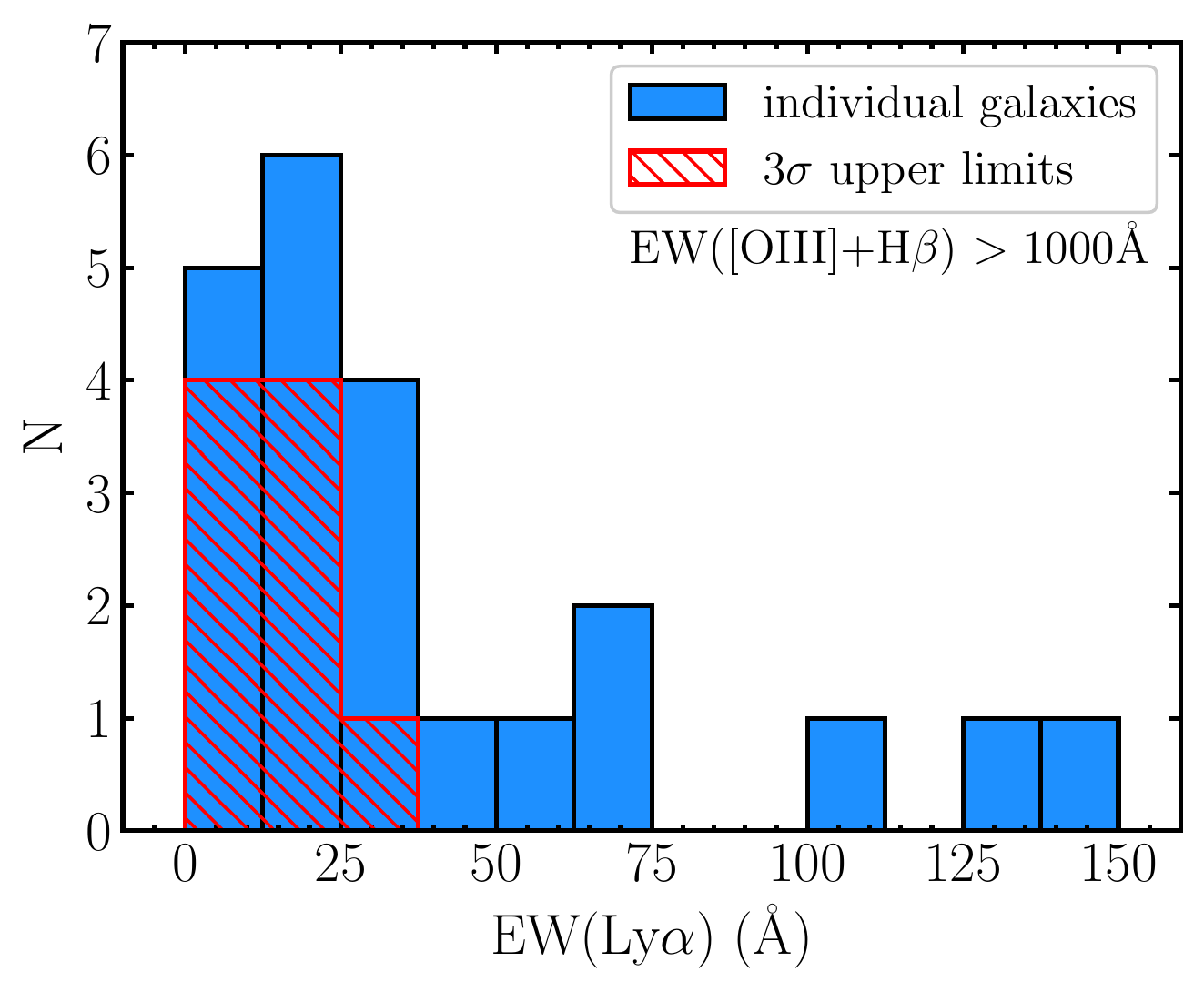}
\caption{Ly$\alpha$ EW distribution of $z\sim2-3$ galaxies with the largest [O~{\scriptsize III}]+H$\beta$ EWs ($>1000$~\AA). The data set shown in this plot combines our spectroscopic sample and the sample in \citet{Du2020}. Sources with Ly$\alpha$ emission line detections are plotted with blue histograms. For those without significant Ly$\alpha$ detections, we plot the $3\sigma$ upper limits with red hatched histograms. Ly$\alpha$ line emission is not uniformly strong in galaxies with intense [O~{\scriptsize III}]+H$\beta$ line emission (EW $>1000$~\AA), we find that $48$ per cent of these systems show Ly$\alpha$ EW below $10-20$~\AA.}
\label{fig:lyaew_dist}
\end{center}
\end{figure}

The second key takeaway from Figure \ref{fig:lyaew_o3hbew} is 
that the Ly$\alpha$ is not uniformly strong among galaxies with 
intense optical nebular line emission (EW$_{\rm{[OIII]+H}\beta}>1000$~\AA).  
We see relatively weak Ly$\alpha$ (EW $<10$~\AA) and 
several non-detections in this population, suggesting significant 
neutral hydrogen covering fractions.
This can be more clearly seen in Figure \ref{fig:lyaew_dist}, 
where we show the Ly$\alpha$ EW distribution of galaxies with 
EW$_{\rm{[OIII]+H}\beta}>1000$~\AA.  This plot shows that 
$48$ per cent of these systems have relatively low Ly$\alpha$ EWs 
($<10-20$~\AA). Thus at least at $z\simeq2-3$, it is evident that not 
all of the extreme [O~{\small III}]+H$\beta$ emitting galaxies are strong 
Ly$\alpha$ emitters. This finding was also reported in \citet{Du2020} 
based on very deep Keck/LRIS spectra (see red open circles in Figure \ref{fig:lyaew_o3hbew}). 
Our survey extends this result to the most extreme [O~{\small III}]+H$\beta$ emitting galaxies. 
Since we expect all systems with intense optical line emission  
(EW$_{\rm{[OIII]+H}\beta}>1000$~\AA) to be efficient producers of 
Ly$\alpha$ (\citealt{Chevallard2018}; \citetalias{Tang2019}), the results 
described above suggest that many of these galaxies have their Ly$\alpha$ 
weakened within the ISM or CGM. If $z>7$ galaxies are similar, we should not expect 
to see strong Ly$\alpha$ in every system with extreme [O~{\small III}]+H$\beta$ 
emission, as has been seen in recent reionization-era surveys \citep{Stark2017,Endsley2021b}.

\begin{figure}
\begin{center}
\includegraphics[width=\linewidth]{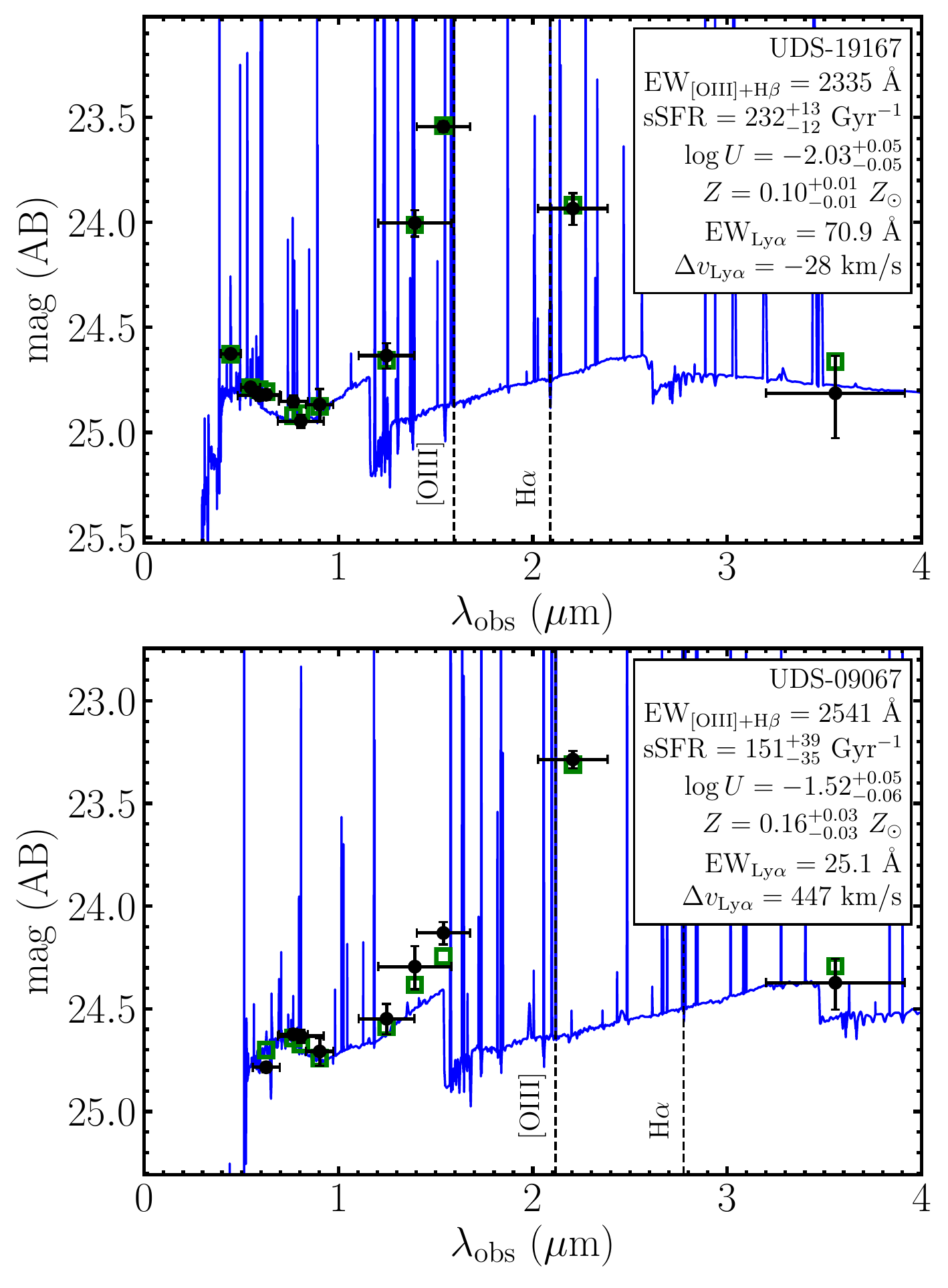}
\caption{Broadband SEDs of two of the most extreme [O~{\scriptsize III}] emitters (EW$_{\rm{[OIII]+H}\beta}>1800$~\AA) with Ly$\alpha$ emission in our sample. The two objects have similar [O~{\scriptsize III}]+H$\beta$ EWs but UDS-19167 (upper panel) shows a larger Ly$\alpha$ EW and smaller velocity offset than UDS-09067 (lower panel). Observed broadband photometry is shown as solid black circles. The best-fit SED models inferred from BEAGLE are plotted by solid blue lines, and synthetic photometry is presented by open green squares. We write the sSFR, the ionization parameter, and the metallicity derived from BEAGLE, as well as the [O~{\scriptsize III}]+H$\beta$ EW, Ly$\alpha$ EW, and velocity offset of each object in the upper right corner.}
\label{fig:lya_sed}
\end{center}
\end{figure}

Our sample allows us to investigate why some extreme [O~{\small III}] emitters  
have strong Ly$\alpha$ emission and others do not. 
Here we consider the seven galaxies with 
the most extreme optical line emission in our sample 
(EW$_{\rm{[OIII]+H}\beta}>1800$~\AA), corresponding to systems 
undergoing a rapid upturn or burst of star formation. 
In this subset, there are four very strong Ly$\alpha$ 
emitters (Ly$\alpha$ EW $>50$~\AA) and three systems with 
weaker or undetected Ly$\alpha$ (see Figure \ref{fig:lya_sed} for two examples). 
According to the best-fit BEAGLE photoionization models (see Section \ref{sec:spec}), 
the four objects with strong Ly$\alpha$ (EW $>50$~\AA) have similarly large sSFRs 
(median sSFR $=239$~Gyr$^{-1}$), large ionization parameters (median $\log{U}=-1.83$), 
and low metallicities (median $Z=0.10\ Z_\odot$) as the three systems with weaker 
Ly$\alpha$ emission (EW $<50$~\AA, median sSFR $=151$~Gyr$^{-1}$, median $\log{U}=-1.52$, 
median $Z=0.16\ Z_\odot$). Thus in our current sample, we do not see 
substantial differences in the stellar and ionized gas properties of 
strong and weak Ly$\alpha$ emitters with EW$_{\rm{[OIII]+H}\beta}>1800$~\AA. 
Both populations appear to be dominated by very young and metal poor stellar populations, 
suggesting broadly similar radiation fields with comparable production 
efficiencies of ionizing (and Ly$\alpha$) photons.  

\begin{figure}
\begin{center}
\includegraphics[width=\linewidth]{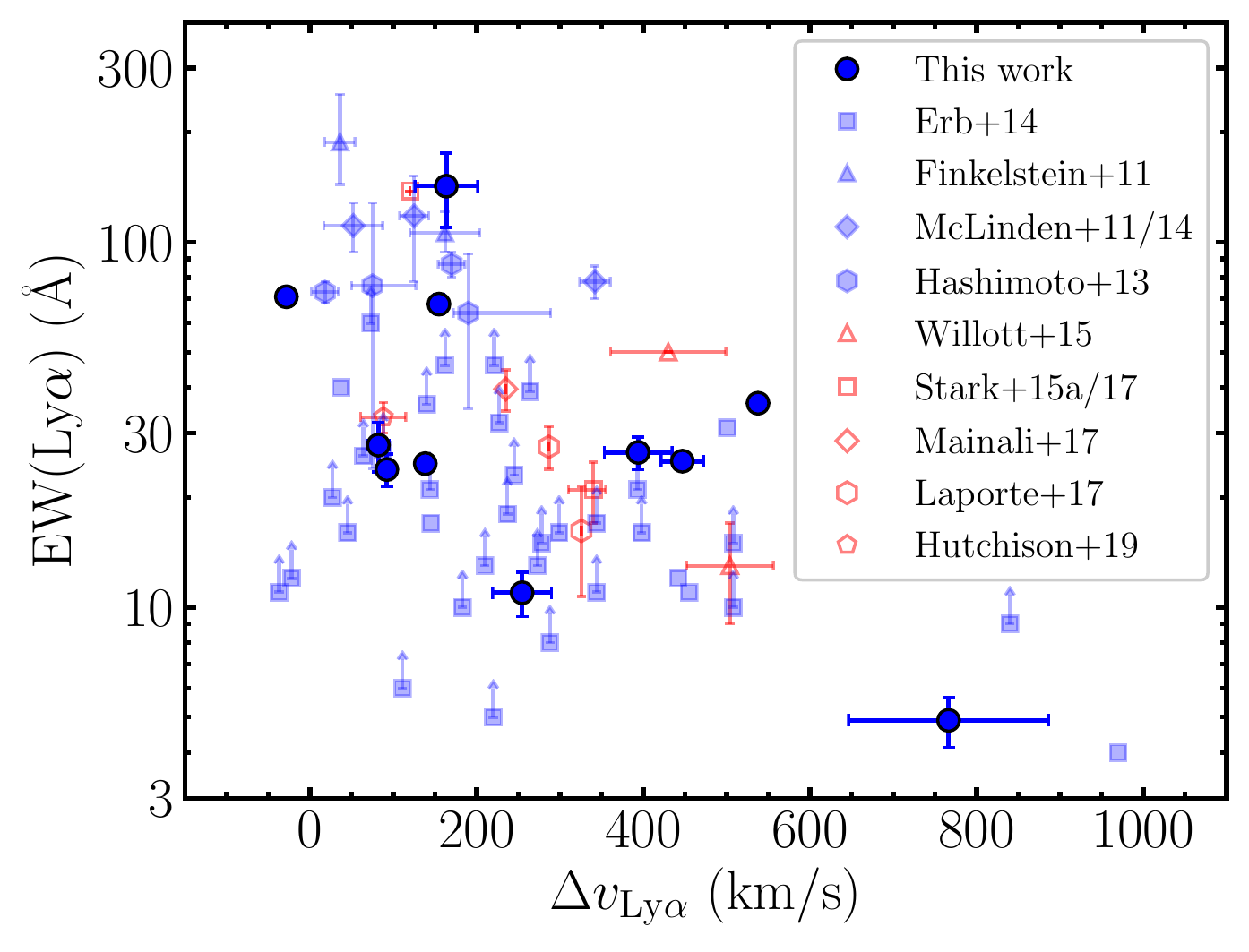}
\caption{The relationship between the Ly$\alpha$ velocity offset and Ly$\alpha$ EW. We present velocity offsets of our sample of extreme [O~{\scriptsize III}] emitters with Ly$\alpha$ emission with blue circles. We also show velocity offsets of $z\sim2-3$ Ly$\alpha$ emitters from literature with blue solid symbols (square: \citealt{Erb2014}; triangle: \citealt{Finkelstein2011}; diamond: \citealt{McLinden2011,McLinden2014}; hexagon: \citealt{Hashimoto2013}), and $z>6$ Ly$\alpha$ emitters with red open symbols (triangle: \citealt{Willott2015}; square: \citealt{Stark2015,Stark2017}; diamond: \citealt{Mainali2017}; hexagon: \citealt{Laporte2017}; pentagon: \citealt{Hutchison2019}). It is clear that galaxies with larger Ly$\alpha$ EWs tend to have smaller Ly$\alpha$ velocity offsets.}
\label{fig:v_lya}
\end{center}
\end{figure}

What does appear different is the velocity offset of Ly$\alpha$ 
with respect to the systemic redshift ($\Delta \rm{v}_{\rm{Ly\alpha}}$). 
Considering only those systems with EW$_{\rm{[OIII]+H}\beta}>1800$~\AA, 
we find that the four galaxies with strong Ly$\alpha$ emission have systematically 
smaller velocity offsets ($\Delta \rm{v}_{\rm{Ly\alpha}}=-28$ km/s 
to $164$ km/s, with a median value of $155$ km/s) 
with respect to the single weaker Ly$\alpha$ emitter where it was possible to measure a 
velocity offset ($\Delta \rm{v}_{\rm{Ly\alpha}}=447$ km/s), a trend that is consistent with 
what is seen in the broader population of star-forming galaxies at these redshifts 
(e.g. \citealt{Finkelstein2011,McLinden2011,McLinden2014,Hashimoto2013,Erb2014}) and with our full 
sample of extreme line emitters with EW$_{\rm{[OIII]+H}\beta}=300-1800$~\AA\  
(Figure \ref{fig:v_lya}). These results may reflect 
some combination of larger column density, covering fraction, 
or velocity dispersion of hydrogen near line center in galaxies 
with weak Ly$\alpha$ emission \citep[e.g.,][]{Erb2014}. As a 
result, Ly$\alpha$ photons are forced to shift significantly in 
wavelength in order to escape. In these galaxies, Ly$\alpha$ photons diffuse spatially 
(often outside of the spectroscopic aperture) and face absorption by dust, both of 
which contribute to the weak Ly$\alpha$ emission. While extreme optical line 
emitters are often associated with strong Ly$\alpha$ emission \citep[e.g.][]{Yang2017a,Stark2017} and 
significant Lyman continuum (LyC) leakage \citep{Izotov2018,Vanzella2016,Vanzella2018}, the
results in Figure \ref{fig:v_lya} indicate that significant hydrogen columns are often 
located in the vicinity of the young super star clusters powering the nebular emission.  

\begin{figure*}
\begin{center}
\includegraphics[width=0.95\linewidth]{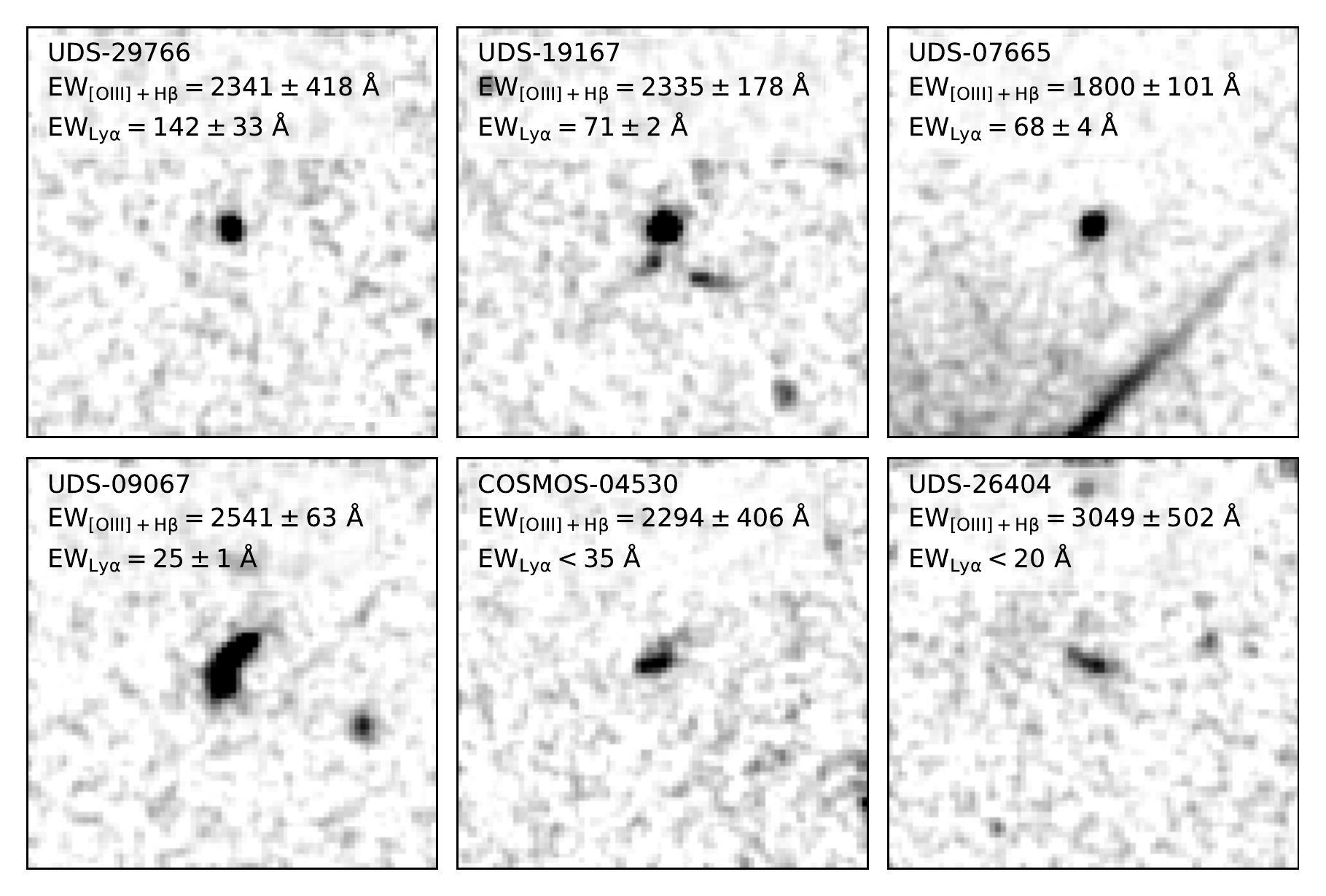}
\caption{{\it HST}/ACS F814W postage stamps ($5''\times5''$ with pixel scale of $0''.06$) of six galaxies with the largest [O~{\scriptsize III}]+H$\beta$ EWs ($>1800$~\AA) in our $z\simeq2-3$ spectroscopic sample. The upper panels show images of three objects with strong Ly$\alpha$ emission (EW $>50$~\AA). These galaxies are characterized by a round shape with low ellipticity ($e=0.11-0.19$). The lower panels show images of other three galaxies with weaker or non-detected Ly$\alpha$ (EW $<50$~\AA). They show irregular or disk-like morphology with much larger ellipticity ($e=0.44-0.64$).}
\label{fig:lya_img}
\end{center}
\end{figure*}

High resolution imaging from {\it HST} highlights another difference between 
strong and weak Ly$\alpha$ emitters in galaxies with extreme optical line 
emission. In Figure \ref{fig:lya_img}, we present color images of six of the 
seven galaxies in our sample with the most intense [O~{\small III}]+H$\beta$ emission  
(EW$_{\rm{[OIII]+H}\beta}>1800$~\AA), suggesting a very recent upturn in 
star formation within the galaxy\footnote{While there are seven galaxies in our 
sample with EW$_{\rm{[OIII]+H}\beta}>1800$~\AA, only six have {\it HST}/ACS imaging.}. 
The three systems in the top row have strong Ly$\alpha$ (EW $=68-142$~\AA) and 
those in the bottom have weak or undetected Ly$\alpha$ (EW $<35$~\AA). 
To quantify the structural parameters of these six objects, we use SExtractor \citep{Bertin1996} 
to measure the half-light radius ($r_{1/2}$) and the ellipticity (defined as $e=1-b/a$, 
where $a$ and $b$ are semi-major and semi-minor axis) from {\it HST}/F814W (rest-frame UV) images. 
We find that the three strong Ly$\alpha$ emitters have slightly smaller radii 
($r_{1/2}=0.49^{+0.01}_{-0.04}$ kpc) comparing to the three galaxies 
with weaker Ly$\alpha$ ($r_{1/2}=0.76^{+0.08}_{-0.09}$ kpc), 
consistent with previous studies indicating that galaxies with larger 
Ly$\alpha$ EWs tend to have smaller sizes \citep[e.g.][]{Law2012,Malhotra2012}. 
We additionally find that strong Ly$\alpha$ emitters have 
lower ellipticities ($e=0.17^{+0.02}_{-0.06}$) than 
those with weaker Ly$\alpha$ ($e=0.64^{+0.00}_{-0.20}$), 
indicating that systems lacking strong Ly$\alpha$ tend to 
have a disk-like or irregular shape. 
This is consistent with results found previously for the general population of 
Ly$\alpha$ emitters at $z\sim2-6$ \citep{Shibuya2014,Kobayashi2016,Paulino-Afonso2018}. 
It has been suggested previously that the range of observed ellipticities may be 
related to the inclination angle of the galaxy \citep{Verhamme2012,Paulino-Afonso2018}. 
In this context, the variation of Ly$\alpha$ EW in the most extreme [O~{\small III}] emitters 
could be explained as an effect of viewing angle, with Ly$\alpha$ photons tending to escape face-on 
(i.e., low ellipticity) following the path of least opacity as suggested by simulations 
\citep[e.g.][]{Verhamme2012,Behrens2014}. However, it is 
not clear that the population of extreme line emitters 
has the same disk-like morphology simulated in these studies, so the inclination 
explanation should be treated with some caution. Regardless of the precise 
explanation, these results suggest that the subset of the extreme [O {\small III}] emitting 
population that appear irregular or disk-like are likely to have sufficient hydrogen 
covering fractions to weaken Ly$\alpha$ emission.  

In the final portion of this section, we now seek to 
provide a baseline for comparison against 
similar measurements in the reionization era. We 
derive the Ly$\alpha$ emitter fraction (x$_{\rm{Ly\alpha}}$) 
as a function of [O~{\small III}]+H$\beta$ EW at $z\simeq2-3$. We 
consider three different [O~{\small III}]+H$\beta$ EW bins 
($300-600$~\AA, $600-900$~\AA, and $900-3000$~\AA). 
To optimize comparison with $z>7$ samples, we only 
consider galaxies with $-21.75<M_{\rm{UV}}<-20.25$. Since 
previous studies show the Ly$\alpha$ fraction strongly depends on 
UV luminosity \citep{Stark2010}, this control will help 
isolate the dependence of Ly$\alpha$ on the 
[O~{\small III}]+H$\beta$ EW. With our $M_{\rm{UV}}$ 
selection applied, we have $9$, $6$, and $10$ objects with EW$_{\rm{[OIII]+H}\beta}=300-600$~\AA,
$=600-900$~\AA, and $=900-3000$~\AA. We compute the fraction of 
galaxies in each bin with Ly$\alpha$ EW $>25$~\AA, including 
both detections and non-detections with robust ($<25$~\AA) upper limits. 
We find that the fraction of galaxies with EW$_{\rm{Ly}\alpha}>25$~\AA\ increases 
with [O~{\small III}]+H$\beta$ EW at $2\sigma$ significance, 
from x$_{\rm{Ly}\alpha}=0.00^{+0.18}_{-0.00}$ 
to $0.17^{+0.29}_{-0.14}$ and $0.40^{+0.20}_{-0.18}$ at 
EW$_{\rm{[OIII]+H}\beta}=300-600$~\AA, $600-900$~\AA, and $900-3000$~\AA. 
We note that the sample size of bright ($-21.75<M_{\rm{UV}}<-20.25$) 
extreme [O~{\small III}] emitters at $z\sim2-3$ is relatively small 
($\lesssim10$ per [O~{\small III}]+H$\beta$ EW bin), 
which is due to the low number density of this population ($\lesssim10$ per $120$ arcmin$^2$). 
In the future, we aim to obtain a larger sample to improve the statistics. 
Since the Ly$\alpha$ fraction closely tracks the UV continuum slope \citep[e.g.][]{Stark2010}, 
we also consider the effects of limiting our measurement to those objects with blue UV slopes 
($\beta<-1.8$) similar to those seen at $z>7$. The same trend 
emerges, albeit with a slightly larger Ly$\alpha$ fraction ($0.50^{+0.22}_{-0.22}$) in 
the bin with largest [O~{\small III}]+H$\beta$ EW.

The results presented above indicate the manner in which Ly$\alpha$ 
EWs increase with [O~{\small III}]+H$\beta$ EWs at $z\simeq2-3$, building on 
results previously presented in \citet{Du2020}. Whether this is 
driven entirely by the increase in the production efficiency of ionizing 
photons (and hence likely the Ly$\alpha$ production efficiency) in extreme optical 
line emitters is not clear. To explore this, we derive the Ly$\alpha$ 
escape fraction as a function of [O~{\small III}]+H$\beta$ EW for the galaxies 
in our sample. The Ly$\alpha$ escape fraction ($f^{\rm{Ly}\alpha}_{\rm{esc}}$) 
is defined as the ratio of the observed Ly$\alpha$ luminosity to 
the intrinsic Ly$\alpha$ luminosity. To compute the intrinsic Ly$\alpha$ luminosity, 
we follow an approach very similar to what has been done previously in the literature 
\citep[e.g.][]{Hayes2010,Erb2014,Henry2015,Trainor2015,Verhamme2017,Yang2017a,Jaskot2019}. 
We assume the Ly$\alpha$/H$\alpha$ flux ratio expected by Case B recombination 
($8.7$; see \citealt{Henry2015} for discussion about the Ly$\alpha$/H$\alpha$ flux ratio) 
and compute the Ly$\alpha$ escape fraction using the following equation: 
$f^{\rm{Ly}\alpha}_{\rm{esc}}=F^{\rm{obs}}_{\rm{Ly}\alpha}/(8.7\times F^{\rm{corrected}}_{\rm{H}\alpha})$. 
For galaxies with H$\alpha$ detections, we use the measured H$\alpha$ fluxes \citepalias{Tang2019}. 
Otherwise we use the H$\alpha$ fluxes inferred from the best-fitting photoionization models. 
To verify that the H$\alpha$ flux predicted by the models is robust, 
we compare the model H$\alpha$ flux and the observed H$\alpha$ flux for 
the subset of galaxies with H$\alpha$ detections. The results reveal good agreement, 
with a median error of only $2.5$ per cent, smaller than the observed uncertainties in 
the H$\alpha$ flux (median uncertainty of $4$ per cent). We perform the dust correction 
to the H$\alpha$ flux assuming the \citet{Calzetti2000} attenuation law, consistent with 
previous studies of Ly$\alpha$ escape fraction \citep[e.g.][]{Hayes2010,Henry2015,Yang2017a}.

For the $21$ galaxies with Ly$\alpha$ detections in our sample, we find that 
the Ly$\alpha$ escape fraction increases with [O~{\small III}]+H$\beta$ EW. 
The median $f^{\rm{Ly}\alpha}_{\rm{esc}}$ increases from $0.02\pm0.01$ at 
EW$_{\rm{[OIII]+H}\beta}=300-600$~\AA\ ($3$ sources) to $0.03\pm0.01$, $0.06\pm0.01$, 
and $0.11\pm0.03$ at EW$_{\rm{[OIII]+H}\beta}=600-900$~\AA\ ($3$ sources), 
$900-1500$~\AA\ ($9$ sources), and $>1500$~\AA\ ($6$ sources) respectively. 
This relationship suggests that the increase of Ly$\alpha$ EW with 
[O~{\small III}]+H$\beta$ EW is not only driven by the increase in the Ly$\alpha$ 
production efficiency, but also the enhanced transmission of Ly$\alpha$ photons 
through the ISM and the CGM in extreme [O~{\small III}] emitters. Physically 
this may indicate that when galaxies go through periods of high sSFR, the feedback 
associated with the recent burst is able to disrupt the surrounding gas sufficiently 
to boost the transfer of Ly$\alpha$ photons. We can also quantify the dependence of 
the Ly$\alpha$ escape fraction on the Ly$\alpha$ EW in our sample. We find that 
the escape fraction increases with Ly$\alpha$ EW, with values of 
$f^{\rm{Ly}\alpha}_{\rm{esc}}\simeq0.02$ at EW$_{\rm{Ly}\alpha}<10$~\AA\ to 
$f^{\rm{Ly}\alpha}_{\rm{esc}}\simeq0.30$ at EW$_{\rm{Ly}\alpha}>100$~\AA. 
The trend we derived here is consistent with the EW$_{\rm{Ly}\alpha}$ vs. 
$f^{\rm{Ly}\alpha}_{\rm{esc}}$ relation inferred from observations of both local and 
high-redshift galaxies \citep[e.g.][]{Verhamme2017,Yang2017a,Jaskot2019}, consistent 
with the picture whereby large Ly$\alpha$ EW traces large Ly$\alpha$ escape fraction. 
We note that in addition to the Ly$\alpha$ production efficiency and the Ly$\alpha$ 
escape fraction, the Ly$\alpha$ EWs are also affected by the absorption of ionizing 
photons by dust in the ionized gas \citep{Charlot2000}. However, because the 
dust attenuation in the extreme emission line galaxies tends not to be significant \citep{Tang2019}, 
this effect is minimal for the galaxies considered here. Indeed our best-fitting photoionization 
models predict that dust absorption of ionizing photons reduces the Balmer lines 
by only $12$ per cent on average. As a result, the increase of Ly$\alpha$ EW 
with [O~{\small III}]+H$\beta$ EW is mainly dominated by the increase of Ly$\alpha$ 
production efficiency and Ly$\alpha$ escape fraction.


\section{Discussion} \label{sec:discussion}

The results described in Section \ref{sec:result} provide a $z\simeq2-3$ framework for 
understanding the Ly$\alpha$ properties of galaxies expected to be typical in the 
reionization era. Here we consider implications for the large Ly$\alpha$ detection rates 
in $z>7$ galaxies with strong [O~{\small III}]+H$\beta$ emission (Section \ref{sec:lya_visibility}) 
and for the ionizing efficiency of this population (Section \ref{sec:escape_fraction}).  

\subsection{Implications for the Ly$\alpha$ visibility at $z>7$} \label{sec:lya_visibility}

The evolving visibility of Ly$\alpha$ emission from star-forming 
galaxies at $z>6.5$ remains one of our primary observational probes 
of the progress of reionization, implying IGM neutral fractions 
in excess of $x_{\rm{HI}}>0.76$ ($68$ per cent confidence) at $z\simeq8$ 
\citep[e.g.][]{Mason2019}. The detection of Ly$\alpha$ in $100$ per cent of 
the galaxies in \citetalias{Roberts-Borsani2016} (each selected to have strong 
[O~{\small III}]+H$\beta$ emission) stands in striking contrast to the 
strong line attenuation experienced by most $z\simeq 7-9$ galaxies.  
Why the \citetalias{Roberts-Borsani2016} objects are detectable in Ly$\alpha$ at 
redshifts where the IGM is thought to be mostly neutral is not clear.  
The Ly$\alpha$ statistics presented in Section \ref{sec:result} provide the baseline 
at $z\simeq2-3$ necessary to understand these results and the 
implications they have for the factors regulating the 
visibility of Ly$\alpha$ in reionization-era galaxies.  

While the optical line EWs of the \citetalias{Roberts-Borsani2016} galaxies are extremely 
large (EW$_{\rm{[OIII]+H\beta}}=900-2000$~\AA; c.f. \citealt{Roberts-Borsani2020}), 
so are those of typical galaxies (EW$_{\rm{[OIII]+H\beta}}=670$~\AA; \citealt{Labbe2013}) 
which generally do not show Ly$\alpha$ at $z>7$. If the detectability of Ly$\alpha$ 
in the \citetalias{Roberts-Borsani2016} galaxies is primarily driven by the radiation field 
associated with the intense [O~{\small III}]+H$\beta$ line emission, it therefore 
suggests a substantial change in the Ly$\alpha$ EW distribution at 
EW$_{\rm{[OIII]+H\beta}}>900$~\AA. Our survey suggests that 
such a trend does indeed exist at $z\simeq2-3$, building on 
previous findings in \citet{Du2020}. Our data indicate that the 
Ly$\alpha$ emitter fraction (EW$_{\rm{Ly\alpha}}>25$~\AA) in 
luminous (M$_{\rm{UV}}<-20.25)$ and blue ($\beta<-1.8$) galaxies 
increases by roughly $3\times$ (at $2\sigma$ significance) between 
[O~{\small III}]+H$\beta$ EW $=600-900$~\AA\ and $900-3000$~\AA. 
In Section \ref{sec:result}, we demonstrated that this trend can be explained 
by a shift toward larger ionizing photon production efficiency and larger 
Ly$\alpha$ escape fractions in galaxies with extreme [O~{\small III}]+H$\beta$ 
emission. These extreme line emitters are those with the largest sSFR 
(Figure \ref{fig:ew_star}), as expected for systems undergoing a burst or 
upturn in star formation. During this presumably brief phase, the Ly$\alpha$ 
emission is enhanced relative to galaxies with lower sSFR.  
Thus by selecting $z\simeq 7-8$ galaxies with the largest 
[O~{\small III}]+H$\beta$ EWs (e.g., \citetalias{Roberts-Borsani2016}), 
one is more likely to select galaxies with Ly$\alpha$ 
emission above current sensitivity limits.  

While such extreme [O~{\small III}]+H$\beta$ emitters are very rare at $z\simeq 2-3$, they 
become increasingly more commonplace in the reionization era \citep{Smit2015,deBarros2019,Endsley2021a}, 
reflecting a shift toward larger sSFRs at earlier times. Indeed, in a given sample 
of $z\simeq 7-8$ galaxies, the [O~{\small III}]+H$\beta$ EWs can be expected 
to span from $300$~\AA\ to $3000$~\AA\ \citep{Endsley2021a}.  As 
can be seen in Figure \ref{fig:lyaew_o3hbew}, this range will show large variations 
in Ly$\alpha$ EW that have nothing to do with the IGM, with 
the most extreme optical line emitters much more likely to show strong Ly$\alpha$ emission. 
The dependence of Ly$\alpha$ on [O~{\small III}]+H$\beta$ EW must be 
considered when using the evolving Ly$\alpha$ properties as a 
probe of reionization. Recent spectroscopic investigations at $z>6.5$ 
have often prioritized sources with large IRAC excesses (and hence 
large [O~{\small III}]+H$\beta$ EW) as these objects have narrow confidence intervals 
on their photometric redshifts, allowing Ly$\alpha$ to be placed 
in regions where atmospheric transmission is large. While this 
increases the likelihood of a meaningful constraint on Ly$\alpha$, 
it also increases the likelihood that Ly$\alpha$ will have an 
atypically large EW, biasing inferences on the Ly$\alpha$ 
EW distribution. These problems can be mitigated in future surveys 
by targeting galaxies with representative values of [O~{\small III}]+H$\beta$ EW, 
while also taking efforts to match galaxies across redshift with 
similar Ly$\alpha$ production efficiencies.

\subsection{Implications for ionizing photon escape from extreme [O~{\small III}] emitters} \label{sec:escape_fraction}

Recent studies have suggested that the extreme optical line emitting 
galaxies may be very effective ionizing agents. Not only do they have large
ionizing production efficiencies (\citealt{Chevallard2018}; \citetalias{Tang2019}), 
but they also may often leak significant fractions of their ionizing radiation into the IGM. 
This latter finding has come to light from rest-frame optical spectra of galaxies 
at $z\simeq0.1-0.3$ and $z\simeq3$ known to be LyC leakers 
\citep[e.g.][]{Izotov2016,Izotov2017,Izotov2018,Fletcher2019,Vanzella2020}. 
In these existing samples, the largest escape fractions 
are commonly associated with very large rest-frame optical line 
equivalent widths ([O~{\small III}]+H$\beta$ EW $>1000-2000$~\AA), indicating 
a population of galaxies that has recently experienced a burst or upturn in 
star formation. These objects also tend to show very large ratios
of their [O~{\small III}] to [O~{\small II}] emission lines (hereafter O32)
\citep{Faisst2016,Izotov2016,Fletcher2019,Vanzella2020}, perhaps indicating 
reduced [O~{\small II}] emission stemming from density-bounded H {\small II} regions 
\citep{Jaskot2013,Nakajima2014} or large ionization parameters 
(see \citealt{Plat2019}). Collectively these observations suggest 
that when galaxies undergo intense bursts of star formation, 
the conditions are often met for LyC leakage. However it has recently 
become clear that intense rest-frame optical nebular emission and large O32 
are not sufficient criteria to guarantee LyC leakage \citep[e.g.][]{Izotov2018,Jaskot2019,Nakajima2020},
potentially indicating that a subset of systems undergoing bursts have significant 
hydrogen columns that completely cover the young stars along the line-of-sight 
(see also \citealt{Katz2020,Barrow2020}). 
The impact of neutral gas on LyC escape can be studied indirectly 
via resonant emission lines (i.e. Ly$\alpha$, Mg~{\small II}) or interstellar 
absorption lines. Galaxies with gas conditions favorable to LyC leakage 
(e.g., low column density, low gas covering fraction) 
show strong Ly$\alpha$ with narrow line profiles 
\citep[e.g.][]{Verhamme2015,Dijkstra2016,Steidel2018,Rivera-Thorsen2019}, 
optically thin Mg~{\small II} emission profiles \citep{Henry2018,Chisholm2020}, and weak interstellar 
absorption lines from low ionization metals \citep{Reddy2016,Steidel2018}. 

As our understanding of the conditions required for LyC leakage 
improves, it so becomes possible to explore whether those 
conditions are met in a large fraction of $z>7$ galaxies. 
The first step toward this goal 
has been realized through characterization of the [O~{\small III}]+H$\beta$ 
strengths at $z\simeq 7$ \citep{Labbe2013,Smit2014}.  
These results indicate that extreme optical 
line emission is much more common at $z\simeq7$ than at lower 
redshifts \citep{deBarros2019,Endsley2021a}. 
{\it JWST} will soon complement these studies 
with measurements of O32. If extreme line emitters at $z\simeq 7$ 
are similar to those at $z\simeq0-3$, we expect the O32 
values to be uniformly large (i.e., O32 $>6-10$) in the subset 
of the population with [O~{\small III}]+H$\beta$ EW in excess of $1000$~\AA\ \citepalias{Tang2019}. 
Taken together, these results suggest that 
a sizeable fraction of the reionization-era population is likely to 
have rest-frame optical spectral properties very similar to many of 
the known LyC leakers at $z\simeq0-3$. But as discussed above, 
large O32 and intense optical line emission do not guarantee 
leakage, as many of these bursts are covered by large enough 
columns of hydrogen to absorb the escaping ionizing radiation.
Ideally Ly$\alpha$ emission line spectra could be used to 
inform the range of line-of-sight neutral hydrogen opacities 
in galaxies at $z\simeq7$ \citep{Matthee2018}, but at 
such high redshifts, these efforts are complicated by the 
impact of the partially-neutral IGM on Ly$\alpha$. So in practice, 
attempts to study Ly$\alpha$ properties in extreme optical 
line emitting galaxies (and implications for LyC escape) 
are best conducted at redshifts after reionization, systematically 
characterizing the statistics of Ly$\alpha$ in galaxies 
matched to the sSFRs that appears common at $z\simeq7$.

The spectra described in this paper allow us to take a step 
in this direction, quantifying the frequency with which $z\simeq 2-3$ extreme emission line galaxies 
have Ly$\alpha$ properties that appear required for LyC leakage. 
These efforts build on studies at $z\simeq0$ \citep{Jaskot2019,Izotov2020} 
and at $z\simeq2-3$ \citep{Du2020}. While our eventual goal is to provide 
a large enough sample to provide a statistical measure of the
Ly$\alpha$ line profiles as a function of rest-frame optical line EWs (or effectively the sSFR), 
here we first consider implications of trends between 
Ly$\alpha$ EW and the [O~{\small III}]+H$\beta$ EW. We are primarily 
interested in galaxies with [O~{\small III}]+H$\beta$ EW $>900$~\AA, 
as these are the systems that have the very large O32 ratios ($>6$; \citetalias{Tang2019}) and 
large star formation rate surface densities that appear frequently linked to 
efficient ionizing photon escape \citep[e.g.][]{Izotov2018,Vanzella2020,Naidu2020}. 
The results described in Section \ref{sec:result} provide two key insights 
into the Ly$\alpha$ properties of this population.  

The spectroscopic sample indicates that very large EW 
Ly$\alpha$ becomes more common in the most extreme optical line 
emitters, consistent with results from nearby galaxies \citep{Yang2017a}.
At high redshift, this was previously hinted at in the 
analysis of \citet{Du2020}. They found that Ly$\alpha$ only becomes 
prominent ($>20$~\AA) at extremely strong [O~{\small III}] emission 
(EW$_{\rm{[OIII]}\lambda\lambda4959,5007}>1000$~\AA, or equivalently EW$_{\rm{[OIII]}\lambda5007}>750$~\AA)
displaying no apparent correlation at lower [O~{\small III}] EWs.
Our sample extends this analysis to higher 
optical line EWs, adding Ly$\alpha$ constraints on 
eleven galaxies with [O~{\small III}]~$\lambda$5007 EW $>1000$~\AA\ to 
the two systems satisfying these criteria in \citet{Du2020}. This 
[O~{\small III}] EW threshold corresponds to [O~{\small III}]+H$\beta$ EW $>1500$~\AA, 
implying a population with extremely large sSFR ($>100$ Gyr$^{-1}$; Figure \ref{fig:ew_star}). 
In this subset, we begin to see extremely strong Ly$\alpha$ emission, 
with some galaxies reaching upwards of Ly$\alpha$ EW $=70-150$~\AA.  
These systems have both efficient Ly$\alpha$ production and 
low enough neutral hydrogen opacities along the line-of-sight to 
facilitate large escape fractions of Ly$\alpha$ (see Section \ref{sec:result}).
Looking at the entire sample with [O~{\small III}]+H$\beta$ EW $>1500$~\AA, 
we find that $50$ per cent have Ly$\alpha$ EW $>25$~\AA, and $38$ per cent have Ly$\alpha$ EW $>50$~\AA, 
both of which are much larger than found in more typical systems at these redshifts. 
These objects appear to be ideal candidates for significant escape fractions, 
with similar rest-frame UV and rest-frame optical spectroscopic 
properties as many of the known LyC leakers. Physically these results 
emphasize the importance of strong bursts (as indicated by extreme
nebular line EWs) in creating the conditions that appear linked to ionizing photon escape.

While Ly$\alpha$ is on average more prominent in galaxies with 
extreme optical line emission, it is not uniformly strong in this population. As is clear from 
above, roughly half of galaxies with sSFR in excess of $100$ Gyr$^{-1}$ have 
weak ($<25$~\AA) Ly$\alpha$ (see Table \ref{tab:lya}). These sources 
tend to have larger Ly$\alpha$ velocity offsets with respect 
to systemic, implying a substantial covering fraction of 
neutral hydrogen at similar velocity as the young star clusters.
This subset of extreme optical line emitters is not likely to leak ionizing radiation 
along the line-of-sight.  From {\it HST} imaging, we see that the 
extreme [O~{\small III}] emitters with weaker Ly$\alpha$ tend to appear 
more disk-like or irregular (Figure \ref{fig:lya_img}). Taken at face value, 
these results suggest that when extreme emission line galaxies 
appear elongated in high resolution imaging, they are more likely 
to have large enough hydrogen covering fractions to reduce the transmission of 
Ly$\alpha$ (and LyC) emission. It is conceivable 
that these objects may be more likely to transmit a larger fraction of 
their Ly$\alpha$ (or LyC) emission if viewed along one of their shorter 
axes. Such viewing angle effects are commonly predicted in simulations 
\citep{Ma2020,Katz2020,Barrow2020} but remain challenging to 
directly confirm observationally.

Overall the results presented here provide continued support for 
indications that the extreme optical line emitting galaxies 
([O~{\small III}]+H$\beta$ EW $>900$~\AA) are very effective ionizing agents.  
While such objects are rare at $z\simeq0-3$, 
they become more common in the $z>7$ population \citep{Smit2014,Smit2015,deBarros2019,Endsley2021a}.  
This reflects an overall shift toward more rapidly rising star formation histories at $z>6$, 
with the systems having the largest sSFRs capable of powering the nebular line emission 
described here. In the future, higher spectral resolution observations 
should be able to characterize the distribution of Ly$\alpha$ line 
profiles as a function of [O~{\small III}]+H$\beta$ EW, providing 
more direct constraints on the likelihood of leaking ionizing radiation 
\citep[e.g.][]{Rivera-Thorsen2017}. Meanwhile, as larger 
samples of extreme [O~{\small III}] emitters lacking Ly$\alpha$ 
are obtained, we should be able to improve our understanding 
of why some systems undergoing rapid upturns in star formation are 
more efficient than others at clearing channels for ionizing photons to escape.


\section{Summary} \label{sec:summary}

We present Ly$\alpha$ equivalent width measurements of $49$ extreme optical line emitting galaxies at $z=2.2-3.7$ with EW$_{\rm{[OIII]+H}\beta}=300-3000$~\AA, similar to the range of optical line EWs seen in reionization-era galaxies and building on previous work presented in \citet{Du2020}. The sample includes $11$ sources with the largest [O~{\small III}]+H$\beta$ EWs ($>1500$~\AA) that characterize many of the known Ly$\alpha$ emitters at $z>7$ (e.g., \citetalias{Roberts-Borsani2016}), enlarging the Ly$\alpha$ statistics for the most extreme [O~{\small III}] emitters at $z\simeq2-3$ by a factor of five. Our data provides an empirical baseline at where the IGM is mostly ionized, allowing us to investigate how factors internal to galaxies impact the Ly$\alpha$ visibility (or lack thereof) in reionization-era galaxies, especially the anomalously large Ly$\alpha$ detection rate of the most extreme [O~{\small III}] line emitting systems at $z>7$ \citep{Stark2017}. We summarize the results below:

(1) We measure the Ly$\alpha$ EW for the $49$ extreme [O~{\small III}] emitters at $z=2.2-3.7$ in our spectroscopic sample. We find that the fraction of strong Ly$\alpha$ emitters (EW$_{\rm{Ly}\alpha}>25$~\AA) scales with the rest-frame optical emission line EW. Considering galaxies in our sample with similar UV luminosities ($-21.75<M_{\rm{UV}}<-20.25$) and blue UV slopes ($\beta<-1.8$) as the $z>7$ objects in \citetalias{Roberts-Borsani2016}, the Ly$\alpha$ emitter fraction (x$_{\rm{Ly}\alpha}=0.50$) of galaxies with EW$_{\rm{[OIII]+H}\beta}>900$~\AA\ (the values probed by \citetalias{Roberts-Borsani2016}) is $\sim3\times$ larger than that (x$_{\rm{Ly}\alpha}=0.20$) of galaxies with EW$_{\rm{[OIII]+H}\beta}$ ($=600-900$~\AA). One of the primary factors driving this trend is the harder radiation field in more intense [O~{\small III}] emitters \citepalias{Tang2019}, leading to larger Ly$\alpha$ production efficiencies. We find that the transmission of Ly$\alpha$ through the ISM and CGM is also likely to increase with EW$_{\rm{[OIII]+H}\beta}$, perhaps reflecting the more intense feedback experienced during the extreme star formation episodes that are associated with large optical line equivalent widths.  

(2) We present the Ly$\alpha$ EW distribution of galaxies with very large [O~{\small III}]+H$\beta$ EWs ($>1000$~\AA) in our sample. Although the fraction of strong Ly$\alpha$ emitter reaches the largest values at these [O~{\small III}]+H$\beta$ EWs, the emerging dataset suggests that $\sim50$ per cent of these systems showing relatively low Ly$\alpha$ EWs ($<10-20$~\AA). Since galaxies with EW$_{\rm{[OIII]+H}\beta}>1000$~\AA\ are found to be very efficient in producing hydrogen ionizing photons (and hence Ly$\alpha$ photons) \citepalias{Tang2019}, the weak Ly$\alpha$ emission likely points to reduced transmission through the ISM and CGM. This result suggests that not all galaxies experiencing a burst or upturn in star formation have cleared pathways allowing Ly$\alpha$ (or LyC) emission to escape.  

(3) To understand why some galaxies undergoing bursts have conditions which facilitate the escape of Ly$\alpha$ and others do not, we explore the properties of galaxies in our sample with the most extreme optical line emission (EW$_{\rm{[OIII]+H}\beta}>1800$~\AA). We find that those systems that are weaker in Ly$\alpha$ tend to have morphologies with larger ellipticities ($e=0.44-0.64$) than those with strong Ly$\alpha$ emission ($e=0.11-0.19$), suggesting that the weak Ly$\alpha$ emitters in this sample of extreme line emitters tend to appear more disk-like or elongated than those with strong Ly$\alpha$ emission. This finding is similar to results seen in the more general population of Ly$\alpha$ emitters \citep{Shibuya2014,Kobayashi2016,Paulino-Afonso2018}. If the ellipticity is set by the observed inclination, extreme line emitters with weak Ly$\alpha$ are most likely to be observed along their longer axis (i.e., edge-on), and those with strong Ly$\alpha$ tend to be seen face-on, similar to predictions from simulations \citep{Verhamme2012,Behrens2014}. These results suggest significant line-of-sight differences in the Ly$\alpha$ opacity through extreme line emitting galaxies. 

(4) We discuss implications of our survey for the findings of \citetalias{Roberts-Borsani2016}, where luminous $z\simeq7-9$ galaxies with extremely large [O~{\small III}]+H$\beta$ EWs are seen with much stronger Ly$\alpha$ emission than the general population at $z>7$ (EW$_{\rm{[OIII]+H}\beta}\sim670$~\AA). For the $z\sim2-3$ sample, the fraction of Ly$\alpha$ emitters (EW $>25$~\AA) among luminous (M$_{\rm{UV}}<-20.25)$ and blue ($\beta<-1.8$) galaxies increases by $3\times$ (at $2\sigma$ significance) from EW$_{\rm{[OIII]+H}\beta}=600-900$~\AA\ to EW$_{\rm{[OIII]+H}\beta}=900-3000$~\AA. This trend can be explained by a shift toward both enhanced ionizing photon (and hence Ly$\alpha$) production efficiency and Ly$\alpha$ escape fraction in galaxies with larger sSFRs (and hence larger [O~{\small III}]+H$\beta$ EWs). These results help explain that by selecting galaxies with the largest [O~{\small III}]+H$\beta$ EWs, one is more likely to select galaxies with detectable large EW Ly$\alpha$ emission.

(5) We discuss the implications for LyC leakage in extreme [O~{\small III}] emitters. Previous work has indicated that this population has uniform large O32 values \citepalias{Tang2019}, similar to those seen in many galaxies with large escape fractions. Overall the results continue supporting the picture that the most extreme optical line emitting galaxies, which become more common at $z>7$, are very effective ionizing agents. Future observations with higher spectral resolution will help to characterize the Ly$\alpha$ emission line profile and provide more direct constraints on LyC leakage.


\section*{Acknowledgements}

We are grateful for enlightening conversations with John Chisholm and Xiaohui Fan.
RE acknowledges funding from JWST/NIRCam contract to the University of Arizona, NAS5-02015. 
EC acknowledges support from ANID project Basal AFB-170002. 
This work is based on observations taken by the 3D-HST Treasury Program (GO 12177 and 12328) with the NASA/ESA HST, 
which is operated by the Association of Universities for Research in Astronomy, Inc., under NASA contract NAS5-26555.
Observations reported here were obtained from the Magellan Telescopes located at Las Campanas Observatory, Chile, 
and the MMT Observatory, a joint facility of the University of Arizona and the Smithsonian Institution. 
This paper uses data products produced by the OIR Telescope Data Center, 
supported by the Smithsonian Astrophysical Observatory.
We acknowledge the MMT queue observers for assisting with MMT/Binospec observations. 

This research made use of {\small ASTROPY}, a community-developed core python package for Astronomy \citep{Astropy2013}, {\small NUMPY}, {\small SCIPY} \citep{Jones2001}, and {\small MATPLOTLIB} \citep{Hunter2007}


\section*{Data Availability}

The data underlying this article will be shared on reasonable request to the corresponding author.



\bibliographystyle{mnras}
\bibliography{Lyman_alpha}



\appendix


\bsp	
\label{lastpage}
\end{document}